\documentclass[12pt]{article}
\usepackage{putex}
\usepackage{graphicx}
\usepackage{epstopdf}
\usepackage{enumerate}
\usepackage{cite}
\usepackage{amsthm}
\usepackage{tikz}
\usetikzlibrary{arrows}
\usetikzlibrary{decorations.markings}

\numberwithin{equation}{section}

\newcommand{\abs}[1]{\left\lvert #1 \right\rvert}

\newcommand {\be} {\begin {equation}}
\newcommand {\ee} {\end {equation}}

\newcommand {\bes} {\begin {equation*}}
\newcommand {\ees} {\end {equation*}}

\newcommand{\es}[2] {\begin{equation} \label{#1} \begin{split} #2 \end{split} \end{equation}}

\newcommand{\Z}{\mathbb{Z}}
\newcommand{\R}{\mathbb{R}}

\newcommand{\C}{\mathbb{C}}

\newcommand{\hx}{\hat{x}}
\newcommand{\hr}{\hat{\rho}}

\def\tr{\operatorname{tr}}
\def\Vol{\operatorname{Vol}}
\def\Area{\operatorname{Area}}
\def\Length{\operatorname{Length}}

\theoremstyle{plain}

\begin{document}

\preprint{PUPT-2384}

\institution{PU}{Joseph Henry Laboratories, Princeton University, Princeton, NJ 08544}

\title{Operator Counting and Eigenvalue Distributions for 3D Supersymmetric Gauge Theories}

\authors{Daniel R.~Gulotta, Christopher P.~Herzog,  and Silviu S.~Pufu}

\abstract{
We give further support for our conjecture relating eigenvalue distributions 
of the Kapustin-Willett-Yaakov matrix model in the large $N$ limit to numbers of operators in the chiral ring of the corresponding supersymmetric three-dimensional gauge theory. 
We show that the relation holds for non-critical R-charges and for examples with ${\mathcal N}=2$ instead of ${\mathcal N}=3$ supersymmetry where the bifundamental matter fields are nonchiral.  We prove that, for non-critical R-charges, the conjecture is equivalent to a relation between the free energy of the gauge theory on a three sphere 
and the volume of a Sasaki manifold that is part of the moduli space of the gauge theory.
We also investigate the consequences of our conjecture for chiral theories where the matrix model is not well understood.
}

\date{June 2011}

\maketitle

\tableofcontents

\newpage

\section{Introduction}

For those interested in superconformal gauge theories in three dimensions, 
the matrix model of Kapustin, Willett, and Yaakov \cite{Kapustin:2009kz} provides a powerful tool.
Using this matrix model, one can compute the partition function and the expectation values of supersymmetric Wilson loops on a three sphere, even when the gauge theory is strongly interacting.  
The matrix model was derived through a localization procedure \cite{Pestun:2007rz} that obscures the connection between matrix model quantities and microscopic degrees of freedom in the gauge theory.   Given the success of the matrix model in post-dicting the $N^{3/2}$ large $N$ scaling of the free energy\footnote{By ``free energy'' we mean minus the logarithm of the path integral on $S^3$, with an appropriate subtraction of UV divergences.} of maximally supersymmetric $SU(N)$ Yang-Mills theory at its infrared fixed point \cite{Drukker:2010nc}, it is a worthwhile goal to try to figure out the map between matrix model and gauge theory quantities in greater detail.
In ref.\ \cite{Gulotta:2011si}, we made some progress in understanding the relation between the eigenvalue distributions in the matrix model and the chiral ring of the supersymmetric gauge theory for the so-called necklace quivers, and we conjectured this relation would hold more generally.  In this paper, we work out further examples of field theories that obey this conjecture.  We restrict ourselves to field theories with M-theory duals of the Freund-Rubin type $AdS_4 \times Y$, where $Y$ is a Sasaki-Einstein space.   A calculation in 11-d supergravity relates the field theory free energy to the volume of the internal space $Y$ on the gravity side through the formula \cite{Herzog:2010hf}
  \es{MtheoryExpectation}{
  F = N^{3/2} \sqrt{\frac{2 \pi^6}{27 \Vol(Y)}} + o(N^{3/2}) \,,
 }
where the normalization of the metric on $Y$ used to compute $\Vol(Y)$ is $R_{mn} = 6 g_{mn}$.

Let us begin by describing the relation noticed in~\cite{Gulotta:2011si} between the eigenvalue distribution in the matrix model and the chiral ring for the necklace quiver gauge theories.   These field theories have ${\mathcal N}=3$ supersymmetry (SUSY), gauge group $U(N)^d$, and associated Chern-Simons levels $k_a$, $a=1, \ldots, d$, such that $\sum_a k_a = 0$.  The matter sector consists of the bifundamental fields $X_{a,a+1}$ and $X_{a+1,a}$ that connect the gauge groups together into a circle (see figure \ref{NecklaceFigure}).  The localization procedure \cite{Kapustin:2009kz} reduces the partition function to an integral over $d$ constant $N\times N$ matrices $\sigma_a$, where $\sigma_a$ is the real scalar that belongs to the same ${\cal N} =2$ multiplet as the gauge connection.  In the large $N$ limit, the matrix integral can be evaluated in the saddle point approximation.  As was shown in \cite{Herzog:2010hf}, at the saddle point, the real parts of the eigenvalues $\lambda^{(a)}_{j}$ of $\sigma_a$ grow as $N^{1/2}$ while their imaginary parts stay of order one as $N$ is taken to infinity.  In addition, to leading order in $N$ the real parts of the eigenvalues are the same for each gauge group.  Therefore, in order to find the saddle point one can consider the large $N$ expansion 
 \es{EvaluesLargeN}{
  \lambda^{(a)}_{ j}  = N^{1/2} x_j + i y_{a,j} + \ldots \,.
 }
As one takes $N \to \infty$, the $x_j$ and $y_{a, j}$ become dense, and one can pass to a continuum description by defining the distributions
 \es{rhoyDef}{
   \rho(x) = \lim_{N \to \infty} \frac{1}{N} \sum_{j=1}^N \delta (x - x_j) \,, \qquad
    \rho(x) y_a(x) = \lim_{N \to \infty} \frac{1}{N} \sum_{j=1}^N y_{a,j} \delta (x - x_j) \,.
 }  
The saddle point is then found by extremizing a free energy functional $F[\rho, y_a]$ under the assumption that $\rho$ is a density, namely that $\rho(x) \geq 0$ and $\int dx\, \rho(x) = 1$.  It is convenient to enforce the latter constraint with a Lagrange multiplier $\mu$ that will appear in the formulae presented below.  In general, $F[\rho, y_a]$ may be a non-local functional because the eigenvalues could interact with one another through long-range forces, and if this is the case the saddle point equations are usually hard to solve.  The key insight in solving the saddle point equations in \cite{Herzog:2010hf} was that, luckily, in the continuum limit \eqref{rhoyDef} the ansatz \eqref{EvaluesLargeN} leads to a {\em local} expression for $F[\rho, y_a]$ due to the cancellation of long-range forces.  By solving the saddle point equations, it was shown in \cite{Herzog:2010hf} that the distributions $\rho(x)$ and $\rho(x) [y_a(x) - y_b(x)]$ can be identified for any $a$ and $b$ with piecewise linear functions with compact support.  While the free energy $F$ can be calculated by evaluating the functional $F[\rho, y_a]$ on the saddle point configuration, it is also possible to calculate $F$ by noticing that $F[\rho, y_a]$ satisfies a virial theorem that gives $F = 4 \pi \mu N^{3/2} / 3$ \cite{Gulotta:2011si}.

The chiral ring of the necklace quiver gauge theories consists of gauge invariant products of the $X_{a,a+1}$ and $X_{a+1,a}$ fields and monopole operators modulo superpotential and monopole relations.  While one can define monopole operators that turn on any number of flux units through each $U(N)$ gauge group, at large $N$ the only relevant ones are the so-called ``diagonal monopole operators'' that turn on the same number of units of flux through the diagonal $U(1)$ subgroup of each $U(N)$ gauge group.  Operators in the chiral ring therefore have an associated R-charge $r$ and a (diagonal) monopole charge $m$.  We can also introduce the the function $\psi_{X_{ab}}(r,m)$ that counts in the same way operators that don't vanish when the bifundamental field $X_{ab}$ is set to zero.\footnote{%
 In our conventions, $X_{ab}$ transforms under the $(\overline {\bf N}_a, {\bf N}_b)$ representation of
 $U(N_a) \times U(N_b)$.
}  
   In \cite{Gulotta:2011si} we found the following relation between the saddle point eigenvalue distribution and the chiral ring:  
\begin{subequations}
\label{results}
\begin{eqnarray}
\label{resultone}
\left. \frac{\partial^3 \psi}{\partial r^2 \partial m} \right|_{m = rx / \mu} &=& \frac{r}{\mu} \rho(x) \ , \\
\label{resulttwo}
\left. \frac{\partial^2 \psi_{X_{ab}}}{\partial r \partial m} \right|_{m=rx / \mu} &=& \frac{r}{\mu} \rho(x)
[ y_b(x) - y_a(x) + R(X_{ab})] \ .
\end{eqnarray}
\end{subequations}
In other words, the matrix model eigenvalue density $\rho(x)$ and the quantity $\rho(x)[ y_b(x) - y_a(x) + R(X_{ab})]$, which as mentioned above are linear functions of $x$, should be interpreted as derivatives of numbers of operators whose monopole charge to R-charge ratio is given by $x/\mu$.

One of the goals of the current paper is to provide further evidence for the conjectures \eqref{results} in superconformal theories with gravity duals that preserve only ${\cal N} = 2$ supersymmetry as opposed to the ${\cal N} = 3$ SUSY of the necklace quivers studied in \cite{Gulotta:2011si}.  In an ${\cal N} = 2$ theory, the $U(1)$ R-symmetry can mix with other Abelian flavor symmetries, so the matter fields can have R-charges different from the canonical free-field value $1/2$.  The generalization of the Kapustin-Willett-Yaakov matrix model to non-canonical R-charges was worked out in \cite{Jafferis:2010un, Hama:2010av}.   Furthermore, since the $U(1)_R$ symmetry can now mix with other Abelian flavor symmetries, it was conjectured in \cite{Jafferis:2010un} that the correct R-symmetry in the IR can be found by extremizing the free energy $F$ as a function of all trial R-charges that are consistent with the marginality of the superpotential.  It has been seen in many examples \cite{Santamaria:2010dm, Jafferis:2011zi, Cheon:2011th, Martelli:2011qj, Amariti:2011hw, Amariti:2011da, Amariti:2011xp, Minwalla:2011ma, Niarchos:2011sn} that this extremum is a maximum and that $F$ is positive.\footnote{It was suggested in \cite{Jafferis:2011zi} that $F$ might be a good measure of the number of degrees of freedom even in non-supersymmetric field theories.  See also \cite{Klebanov:2011gs}.}

We find that eqs.~\eqref{results} are satisfied for more general quiver gauge theories where the bifundamental matter multiplets are non-chiral, meaning that they come in pairs of conjugate representations of the gauge group.  
In the first half of section \ref{sec:centralexamples}, we examine the necklace quiver gauge theories, this time with an arbitrary R-charge assignment consistent with the marginality of the superpotential.  
In the second half of section \ref{sec:centralexamples} and appendix \ref{app:furtherexamples}, we 
examine theories where we add flavor (meaning ${\cal N} = 2$ matter multiplets that transform in the fundamental or anti-fundamental representation of one of the gauge groups)  to the maximally SUSY ${\mathcal N}=8$ theory and to the ${\cal N} = 6$ ABJM theory of  \cite{Aharony:2008ug}.  Lastly, in appendix \ref{sec:c3z2z2}, we consider a theory that shares the same quiver with its $(3+1)$-dimensional cousin that has a $\mathbb{C}^3/\mathbb{Z}_2 \times \mathbb{Z}_2$ moduli space (see figure \ref{C3Z2Z2Figure}).  In all of these examples, eqs.~\eqref{results} are satisfied for any choice of trial R-charges.

Another goal of this paper is to relate the conjecture \eqref{results} to the observation made in \cite{Martelli:2011qj,Jafferis:2011zi} that, as checked in a number of examples, the relation \eqref{MtheoryExpectation} between the free energy and the volume of the internal space holds for any trial R-charges, and not just the ones that extremize $F$.  That this relation\footnote{%
 A similar relation between the anomaly coefficient $a$ computed with a set of trial 
 R-charges and the volume of a 5-d Sasakian space $Y$ is known to hold 
 in theories with $AdS_5$ duals \cite{Butti:2005ps, Eager:2010yu}.
 }  
holds for any trial R-charges is surprising because only for 
the critical R-charges does there exist a known 11-d 
supergravity background $AdS_4 \times Y$.
For non-critical R-charges, measured geometrically in terms of the volume of some corresponding five-cycles of $Y$, one can still identify a class of Sasakian metrics on $Y$ and compute their volume.  The volume $\Vol(Y)$ is a function of the Reeb vector of $Y$, which parameterizes the way the $U(1)_R$ symmetry sits within the isometry group of $Y$.  We show 
in section \ref{sec:proofs}
that \eqref{resultone} holds for some choice of trial R-charges if and only if eq.~\eqref{MtheoryExpectation} holds for the same choice of trial R-charges for the matter fields and a range of R-charges for the monopole operators.
We also show an analogous result that relates \eqref{resulttwo} to the volumes of five-dimensional sub-manifolds of
$Y$.  
For a gauge invariant operator constructed from a closed loop of bifundamental fields $X_{ab}$, it must be true that $\sum_{X_{ab}} (y_a - y_b) = 0$.  Given (\ref{results}), there is a geometric version of this sum that must also vanish.  The last part of section \ref{sec:proofs} explains why.

There are previously recognized difficulties, involving cancellation of long-range forces, in using the matrix model to study the large $N$ limit of theories with chiral bifundamental fields \cite{Jafferis:2011zi}.  We do not surmount these difficulties, but we investigate in section \ref{sec:c3z3} what (\ref{resultone}) and (\ref{resulttwo}) predict for a theory with a moduli space that is a fibration over $\mathbb{C}^3 / \mathbb{Z}_3$ (see figure~\ref{C3Z3Figure}).  We also study a field theory that was conjectured to be dual to $AdS_4 \times Q^{2, 2, 2} / \Z_k$ in appendix \ref{app:Q222} (see figure \ref{Q222Figure}).

The paper contains two heretofore unmentioned appendices.
Appendix \ref{FMAXIMIZATIONNECKLACE} proves that the critical R-charges maximize $F$ for the necklace quivers.
Appendix \ref{app:toric} reviews how to count gauge invariant operators for an Abelian gauge theory with a toric branch of its moduli space.

\section{Matrix models at non-critical R-charges}
\label{NonCriticalR}

\subsection{Review of the large $N$ limit}

To understand what it means to consider non-canonical (or non-critical) R-charges,  
let us introduce some of the ideas developed recently in refs.~\cite{Jafferis:2010un, Hama:2010av, Jafferis:2011zi}.  Building on the work of \cite{Kapustin:2009kz}, refs.~\cite{Jafferis:2010un, Hama:2010av} used localization to reduce the path integral of any ${\cal N} = 2$ Chern-Simons matter on $S^3$ to a matrix integral.  By a Chern-Simons-matter theory we mean a theory constructed from some number $d$ of ${\cal N}= 2$ vector multiplets with gauge groups $G_a$ ($a = 1, \ldots, d$) and Chern-Simons kinetic terms $i \pi k_a \int \tr A_a \wedge dA_a + \text{supersymmetric completion}$, as well as any number of ${\cal N}=2$ chiral superfields transforming in representations $R_i$ of the total gauge group $G = \prod_{a=1}^d G_a$.   As mentioned in the introduction, one difference between theories with ${\cal N}=2$ supersymmetry and theories with more supersymmetry is that the R-charges $\Delta_i$ of the chiral fields at the IR superconformal fixed point are not fixed at the free field values $\Delta_i = 1/2$, so the free energy will generically depend on these R-charges.  In fact, it was proposed in \cite{Jafferis:2010un} that a prescription for finding the correct R-charges in the IR is to calculate the free energy $F$ as a function of all possible R-charge assignments consistent with the marginality of the superpotential and to extremize $F$ over the set of all such assignments.

Let us focus on the case where all gauge groups are $U(N)$ and index the gauge groups by $a = 1, \ldots, d$.   Generalizing the techniques developed in \cite{Herzog:2010hf}, the authors of \cite{Jafferis:2011zi} used the saddle point approximation to evaluate the path integral on $S^3$ for a class of ${\cal N} = 2$ Chern-Simons-matter theories at large $N$ that satisfy the following five conditions:
 \begin{enumerate}
  \item The CS levels sum to zero: $\sum_{a=1}^d k_a = 0$.
  \item Any matter field $X$ transforms either in the ${\bf N}_a$, or $\overline{\bf N}_b$, or $({\bf N}_a, \overline{\bf N}_b)$ representation for some $a$ and $b$.
  \item The total number of fundamental fields equals the total number of anti-fundamental fields.
  \item For any bifundamental field $X$ transforming in $({\bf N}_a, \overline{\bf N}_b)$, there exists another bifundamental field $\tilde X$ transforming in the conjugate representation $({\bf N}_b, \overline{\bf N}_a)$.
  \item For each gauge group $a$ we have
   \es{BetaCond}{
    \sum_\text{$X$ in $({\bf N}_a, \overline{\bf N}_b)$} \left( R[X] - 1 \right)
     + \sum_\text{$\tilde X$ in $({\bf N}_b, \overline{\bf N}_a)$} \left( R[\tilde X] - 1 \right) = -2 \,.
   }
 \end{enumerate}
 This last condition is sufficient to guarantee the vanishing of the long-range forces on the eigenvalues in the saddle point approximation.  
Interestingly, this condition has appeared before in the context of superconformal $(3+1)$-dimensional gauge theories.  
The condition (\ref{BetaCond}) would imply that the NSVZ beta function of gauge group $a$ vanishes \cite{Martelli:2011qj}.  
For quiver gauge theories with a toric moduli space, bifundamental fields appear in exactly two terms in the superpotential.  Thus, if we sum (\ref{BetaCond}) over $a$, we find the condition that 
 \es{Torus}{
(\# \mbox{ of gauge groups})- (\# \mbox{ of bifundamentals})+ (\# \mbox{ of superpotential terms}) = 0 \ .
 }
In other words the quiver may give a triangulation of a torus where the faces of the triangulation are superpotential terms \cite{Franco:2005rj}.

If these five conditions are satisfied, one can take the $N \to \infty$ limit as described in eqs.~\eqref{EvaluesLargeN} and \eqref{rhoyDef} in the introduction.  The free energy is the extremum of the free energy functional
 \es{FFunctional}{
  &F[\rho(x), y_a(x)] = 2 \pi N^{3/2} \int dx\, x \rho(x)  \left( \sum_{a=1}^d k_a y_a(x) + \Delta_m \right) \\
   {}&+ 2 \pi N^{3/2} \int dx\, \abs{x} \rho(x) \left[ \sum_\text{$X$ in ${\bf N}_a$} \left(\frac{1 - R[X]}{2} - \frac{1}{2} y_a(x) \right)
   + \sum_\text{$X$ in $\overline{\bf N}_b$} \left(\frac{1 - R[X]}{2} + \frac{1}{2} y_b(x) \right)   \right]\\
   {}&+ \frac{\pi N^{3/2}}{3}  \int dx\, \rho(x)^2 \sum_\text{$X$ in $({\bf N}_a, \overline{\bf N}_b)$}
    (\delta y_{ab}(x) + R[X]) (\delta y_{ab}(x) + R[X] - 1) (\delta y_{ab}(x) + R[X] - 2)\,,
 }
where $\delta y_{ab}(x) \equiv y_a(x) - y_b(x)$.   This formula was derived assuming the bifundamental fields satisfy $0 \leq R[X] + \delta y_{ab}(x) \leq 2$.  Extra care must be taken when $R[X]  + \delta y_{ab} = 0$ or 2 because in these cases the discrete nature of the eigenvalues becomes important, and the equation of motion derived from varying \eqref{FFunctional} with respect to $\delta y_{ab}(x)$ need not hold.

Generically, the functional~\ref{FFunctional} has many flat directions.  The following $d$ of them play an important role in this paper because they correspond to changing the R-charges of the matter fields by linear combinations of the gauge charges with respect to the diagonal $U(1) \in U(N)_a$:  
 \es{Symmetries}{
  \text{$y_a(x)$:} \qquad &
  y_a(x) \to y_a(x) - \delta^{(a)} \,, \\
  \text{chiral superfield $X$ in ${\bf N}_a$:} \qquad &
   R[X] \to R[X] + \delta^{(a)} \,, \\
  \text{chiral superfield $X$ in $\overline{\bf N}_b$:} \qquad &
   R[X] \to R[X] - \delta^{(b)} \,, \\   
  \text{chiral superfield $X$ in $({\bf N}_a, \overline{\bf N}_b)$:} \qquad &
   R[X] \to R[X] + \delta^{(a)} - \delta^{(b)} \,, \\      
   \text{$\Delta_m$:} \qquad &
   \Delta_m \to \Delta_m + \sum_a k_a \delta^{(a)} \,.
 }
See \cite{Jafferis:2011zi} for a more detailed discussion of these flat directions and their AdS/CFT interpretation.

The $\Delta_m$ appearing in \eqref{FFunctional} is the bare R-charge of the ``diagonal'' monopole
operator $T^{(1)}$.  A monopole operator $T_a^{(q_a)}$ 
turns on $q_a$ units of $\tr F_a$ flux through a two-sphere surrounding the insertion point. 
Diagonal monopole operators $T^{(m)}$ turn on the same number $m$ of 
$\tr F_a$ flux units in each gauge group. At large $N$, only the diagonal monopole operators
are important.

We will usually impose the constraint $\int dx\,\rho(x)=1$ by introducing a Lagrange multiplier $\mu$ and defining the functional
 \es{Ftilde}{
  \tilde F[\rho, y_a, \mu] = F[\rho, y_a] - 2 \pi N^{3/2} \mu \left(\int dx\, \rho(x) - 1 \right) \,.
 }
This functional should be extremized with respect to $\rho(x)$, $y_a(x)$, and $\mu$.

\subsection{Flavored theories}

In all gauge theories that we examine in this paper the fundamental and anti-fundamental fields $q_\alpha$ and $\tilde q_\alpha$ appear in the superpotential as
 \es{FundSuperpot}{
  \delta W = \sum_\alpha \tr \left[ q_\alpha {\cal O}_\alpha \tilde q_\alpha \right] \,,
 }
where ${\cal O}_\alpha$ are polynomials in the bifundamental fields.  It was conjectured in \cite{Benini:2009qs, Jafferis:2009th} that if this is case then the diagonal monopole operators $T^{(m)}$ satisfy the following OPE:
 \es{TTtildeOPE}{
  T^{(m)} T^{(-m)} = \left( \prod_\alpha {\cal O}_\alpha \right)^{\abs{m}} \,.
 }
This OPE was conjectured in part because a parity anomaly argument shows that the monopole operators have gauge charges
 \es{TGauge}{
  g_a[T^{(m)}] = m k_a + \frac{\abs{m}}{2} \sum_\alpha g_a[{\cal O}_\alpha] 
 }
with respect to the diagonal $U(1) \subset U(N)_a$, and R-charges
 \es{TRcharge}{
 R[T^{(m)}] = m \Delta_m + \frac{\abs{m}}{2} \sum_\alpha R[{\cal O}_\alpha] \,.
 }
Using the fact that each term in \eqref{FundSuperpot} must be gauge-invariant and have R-charge two, we have $ R[q_\alpha] + R[\tilde q_\alpha] + R[{\cal O}_\alpha] = 2$ and $g_a[q_\alpha] + g_a[\tilde q_\alpha] + g_a[{\cal O}_\alpha] = 0 $ for any $a$.  One can use these relations to eliminate the sum over the flavor fields in \eqref{FFunctional}:
 \es{FFunctionalAgain}{
  &F[\rho(x), y_a(x)] = 2 \pi N^{3/2} \int dx\, \abs{x} \rho(x)\left(R[T^{(\sgn x)}] +  \sum_{a=1}^d  y_a(x) g_a[T^{(\sgn x)}] \right) \\
   {}&+ \frac{\pi N^{3/2}}{3}  \int dx\, \rho(x)^2 \sum_\text{$X$ in $({\bf N}_a, \overline{\bf N}_b)$}
    (\delta y_{ab}(x) + R[X]) (\delta y_{ab}(x) + R[X] - 1) (\delta y_{ab}(x) + R[X] - 2)\,.
 }

\section{An Equivalent form of our conjecture}
\label{sec:proofs}

\subsection{Eigenvalue density and volumes of Sasakian spaces}
\label{EQUIVRHO}

We can relate the conjecture \eqref{results} to the observation that eq.~\eqref{MtheoryExpectation} holds for any trial R-charges.  In particular, we prove the following result:   In a CS-matter theory dual to $AdS_4 \times Y$ fix a set of matter R-charges $R[X]$ and a bare monopole charge $\Delta_m$ so that the conformal dimensions of all gauge-invariant operators satisfy the unitarity bound.   Let $\rho(x)$, $\mu$, and $\psi(r, m)$ be as defined in the introduction, and let's assume $\rho(x)$ has compact support.  The following two statements are equivalent:
   \begin{enumerate}[A.]
    \item The conjecture~\eqref{resultone} holds for the given R-charges $R[X]$ and bare monopole charge $\Delta_m$. \label{FirstAssertion}
    \item  For any $\delta$ in a small enough neighborhood of zero, we have
     \es{VolYConjecture}{
      \lim_{N \to \infty} \frac{2 \pi^6 N^3}{27 F^2} = \Vol(Y, \delta) \,,
     }
    where the free energy $F$ of the CS-matter theory and the volume $\Vol(Y, \delta)$ of the internal space $Y$ are both computed assuming that the matter R-charges are $R[X]$ and the bare monopole charge is $\Delta_m + \delta$. \label{SecondAssertion}
   \end{enumerate}

For notational convenience, let's denote the LHS of eq.~\eqref{VolYConjecture} by $\Vol_m(Y, \delta)$ and let's introduce the rescaled matrix model quantities:
 \es{hatDefs}{
  \hat x = \frac{x}{\mu} \,, \qquad \hat \rho(\hat x) = \frac{\rho(x)}{\mu} \,, \qquad
   \hat y_a(\hat x) = y_a (x) \,.
 }
The equivalence between \eqref{FirstAssertion} and \eqref{SecondAssertion} follows from the following two equations:
 \es{VolY7MatrixLemma}{
  \Vol_m(Y, \delta) 
   = \frac{\pi^4}{24} \int d\hat x \, \frac{\hat \rho(\hat x)}{\left(1 + \hat x \delta \right)^3} \,,
 }
 \es{VolY7CountingLemma}{
  \Vol(Y, \delta) 
   =\frac{\pi^4}{24} \int d\hat x \, \frac{ \lim_{r \to \infty}  \psi^{(2,1)}(r,r\hx)/r}{\left(1 + \hat x \delta \right)^3} \,,
 }
which we prove in sections \ref{DeltaMDependence} and \ref{DeltaMGeometry}, respectively.

Assuming the eqs.~\eqref{VolY7MatrixLemma} and \eqref{VolY7CountingLemma} to be true, it is clear that the statement \eqref{FirstAssertion} implies \eqref{SecondAssertion}.  That \eqref{SecondAssertion} implies \eqref{FirstAssertion} follows from the fact that knowing $\Vol_m(Y, \delta)$ for $\delta$ in a small neighborhood of zero, one can reconstruct $\hat \rho(\hat x)$, and analogously, from $\Vol(Y, \delta)$ one can reconstruct $\lim_{r \to \infty}  \psi^{(2,1)}(r,r\hx)/r$.  Indeed, one can extend $\Vol_m(Y, \delta)$ to any complex $\delta$ as an analytic function with singularities.  We assume that $\hat \rho$ is supported on $[\hat x_-, \hat x_+]$ for some $\hat x_- < 0 < \hat x_+$.   We see from eq.~\eqref{VolY7MatrixLemma} that the integral converges absolutely if $\delta \in ( - 1/\hat x_+, -1/\hat x_-)$ or $\delta \not \in \R$, so $\Vol_m(Y, \delta)$ can only have singularities on $(-\infty, -1/\hat x_+] \cup [-1/\hat x_-, \infty)$. 
 
To relate the singularities of $\Vol_m(Y, \delta)$ to $\hat \rho(\hat x)$ we can perform two integrations by parts in \eqref{VolY7Delta}
 \es{VolY7DeltaAgain}{
  \Vol_m(Y, \delta) = \frac{\pi^4}{48 \delta^3} \int d\hat x\, \frac{\hat \rho''(\hat x)}{\hat x + \frac{1}{\delta}} 
 }
for any $\delta \in \C \backslash \left((-\infty, -1/\hat x_+] \cup [-1/\hat x_-, \infty)\right)$.  Generically, eq.~\eqref{VolY7DeltaAgain} shows that $\Vol_m(Y, \delta)$ has two branch cuts, one on $(-\infty, -1/\hat x_+]$ and one on $ [-1/\hat x_-, \infty)$.  From the discontinuities of $\Vol_m(Y, \delta)$ one can read off $\hat \rho''(-1/\delta)$.  Simple poles of $\Vol_m(Y, \delta)$ at $\delta = -1/\hat x'$ correspond to contributions to $\hat \rho''(\hat x)$ proportional to $\delta(\hat x - \hat x')$; second order poles of $\Vol_m(Y, \delta)$ at $\delta = -1 / \hat x'$ correspond to $\delta'(\hat x - \hat x')$, etc.  From the singularities of the analytic continuation of $\Vol_m(Y, \delta)$ one can therefore reconstruct uniquely $\hat \rho''(\hat x)$, and hence $\hat \rho(\hat x)$, and similarly for $\Vol(Y, \delta)$ and $\lim_{r \to \infty}  \psi^{(2,1)}(r,r\hx)/r$.  If $\Vol_m(Y, \delta)$ and $\Vol(Y, \delta)$ agree on an open set, then \eqref{FirstAssertion} holds.

In our examples, $\Vol_m(Y, \delta)$ is a rational function of $\delta$ with poles of order at most three, so $\hat \rho (\hat x)$ is piecewise linear and it may have delta-functions.   From the location and residues of the poles one can first reconstruct $\hat \rho''(\hat x)$, and then $\hat \rho(\hat x)$ by integrating $\hat \rho''(\hat x)$ twice.  To perform this reconstruction starting with $\Vol_m(Y, \delta)$, one first decomposes $\Vol_m(Y, \delta)$ into partial fractions, and then identifies the terms in $\hat \rho''(\hat x)$ that give those partial fractions:  if, for example,
 \es{VolAnsatz}{
  \Vol_m(Y, \delta) = \frac{\pi^4}{48 \delta^2} \sum_i \frac{a_i}{1 + \hat x_i \delta} - \frac{\pi^4}{48 \delta} \sum_i 
   \frac{b_i}{(1 + \hat x_i \delta)^2} 
 }
for some $\hat x_i$, then
 \es{rhoAnsatz}{
  \hat \rho''(\hat x) = \sum_i a_i \delta(\hat x - \hat x_i) + \sum_i b_i \delta'(\hat x - \hat x_i) \,.
 }

\subsection{Matrix model dependence on $\delta$}
\label{DeltaMDependence}

In this subsection we prove the result \eqref{VolY7MatrixLemma}.  As we have seen in the previous section, the matrix model generally takes the form
 \es{FreeFunctional}{
  \tilde F[\rho, y_a, \mu] &= \int dx\, \rho(x)^2 f(y_a(x)) - \int dx\, \rho(x) V(x, y_a(x)) \\
  {}&+ 2 \pi N^{3/2} \int dx\, \abs{x} \rho(x) R[T^{(\sgn x)}] - 2 \pi N^{3/2} \mu \left(\int dx\, \rho(x) - 1 \right)  \,,
 }
for some functions $f$ and $V$.   While the explicit form of these function is given in \eqref{FFunctionalAgain}, their precise form doesn't matter.  The only property of $V$ that we will use is that it is homogeneous of degree one in $x$, namely $V(\lambda x, y_a(x)) = \lambda V(x, y_a(x))$ for any $\lambda > 0$.   With the rescaling~\eqref{hatDefs}, one can write $\tilde F$ as 
 \es{tildeFRewrite}{
  \tilde F[\hat \rho, \hat y_a, \mu] = -2 \pi N^{3/2} \mu + \mu^3 \int d\hat x\, \hat x^2 \biggl[ \frac{\hat \rho(\hat x)^2}{\hat x^2} 
    f(\hat y_a(\hat x)) - \frac{\hat \rho(\hat x)}{\hat x} \frac{V(\hat x, \hat y_a(\hat x))}{\hat x}
   \\+ 2 \pi N^{3/2} \frac{\hat \rho(\hat x)}{\abs{\hat x}} \left(R[T^{(\sgn \hat x)}]  - \frac{1}{\abs{\hat x}} \right)
   \biggr] \,.
 }

The rescaling \eqref{hatDefs} is useful because now the equations of motion for $\hat \rho$ and $\hat y_a$ do not involve $\mu$.  One can first solve these equations, and then $\mu$ can be found by integrating $\hat \rho$:  the normalization condition $\int dx \, \rho(x) = 1$ becomes
 \es{Normalizationrhohat}{
  \int d\hat x\, \hat \rho(\hat x) = \frac{1}{\mu^2} \,.
 }  

We now see that if we extremized \eqref{tildeFRewrite} in the case where the monople R-charges were $R[T^{(\pm 1)}]$, we could obtain the saddle point when they are $R[T^{(\pm 1)}] \pm \delta^{(\pm 1)}$ through the transformation:
 \es{NontrivialR}{
  \frac{\hat \rho_\delta(\hat x_\delta)}{ {\hat x_\delta}} &= \frac{\hat \rho(\hat x)}{\hat x} \,, \qquad \qquad
   \frac{1}{\hat x_\delta} = \frac{1}{\hat x} + \delta^{(\sgn \hat x)} \,, \\
  \hat y_{a,\delta}(\hat x_\delta) &= \hat y_a (\hat x)\,, \qquad
   R[T^{(\pm 1)}_\delta] = R[T^{(\pm 1)}] \pm \delta^{(\pm 1)} \,.
 }
Indeed, the equations of motion for $\hat \rho$ and $\hat y_a$ are obtained by extremizing the expression in the square brackets in \eqref{tildeFRewrite}, and this expression is invariant under \eqref{NontrivialR}.  Given that $\hat \rho$ has compact support, the transformations \eqref{NontrivialR} make sense only when $\delta^{(\pm)}$ are small enough.

For simplicity, from now on let's restrict ourselves to the case $\delta^{(+1)} = \delta^{(-1)} = \delta$, even though one can make similar arguments for the case where $\delta^{(+1)}$ and $\delta^{(-1)}$ are arbitrary or satisfy a different relation.   In \cite{Gulotta:2011si}, we showed that $F = 4\pi N^{3/2}\mu/3$, which implies that
 \es{VolY7Delta}{
  \Vol_m(Y, \delta) = \frac{\pi^4}{24 \mu_\delta^2} = \frac{\pi^4}{24} \int d \hat x_\delta \, \hat \rho_\delta(\hat x_\delta)
   = \frac{\pi^4}{24} \int d\hat x \, \frac{\hat \rho(\hat x)}{\left(1 + \hat x \delta \right)^3} \,.
 }

\subsection{Operator counting dependence on $\delta$}
\label{DeltaMGeometry}

We now prove the result \eqref{VolY7CountingLemma}.  Let $A$ be the chiral ring associated to the superconformal field theory dual to $AdS_4 \times Y$ in the Abelian case $N=1$.  $A$ is also a vector space over $\C$ that is graded by the R-charge and monopole charge, meaning that one can define a basis of operators with well-defined R-charge and monopole charge.   Let $A_{m,r}$ be the vector subspace of elements of $A$ with monopole charge $m$ and R-charge $r$.
We introduce the Hilbert-Poincar\'e series
 \es{HPSeries}{
   f(t,u) = \sum_{m,r} \dim (A_{m,r}) t^{r} u^{m} \,.
 }
Since the Abelian moduli space of the gauge theory is the Calabi-Yau cone over $Y$ one can view the operators in the chiral ring as holomorphic functions on this cone.  Martelli, Sparks, and Yau \cite{Martelli:2006yb} show that
 \es{HilbertY}{
  \Vol(Y,\delta) = \frac{\pi^4}{48} \lim_{t \to 1} (1-t)^4 f(t,t^{\delta}) \,.
 }

One can compute the Hilbert-Poincar\'e series for $Y$ in terms of $\psi(r,m)$, the number of operators with R-charge at most $r$ and monopole charge at most $m$.  Approximating $\psi$ by a continuous function of homogeneous degree four, the definition \eqref{HPSeries} gives
 \es{HilbertYpsi}{
  f(t, u) \approx \int dr\, dm\, \psi^{(1, 1)}(r, m) t^r u^m \,.
 }
Since $1-t \approx -\ln t$ for $t \approx 1$, we can use \eqref{HilbertY} and \eqref{HilbertYpsi} to write $\Vol(Y, \delta)$ as
 \es{VolY7Hilbert}{
\Vol(Y,\delta) &\approx \frac{\pi^4}{48} (\ln t)^{4} \int dr\,dm\,\psi^{(1,1)}(r,m) t^{r+m \delta} \\
&= -\frac{\pi^4}{48} (\ln t)^{3} \int dr\,dm\,\psi^{(2,1)}(r,m) t^{r+m \delta} \\
&= -\frac{\pi^4}{48} (\ln t)^{3} \int dr\,d\hx\,r \psi^{(2,1)}(r,r\hx) t^{r(1+\hx \delta)} \\
&= \frac{\pi^4}{24} \int d\hat x\, \frac{\psi^{(2,1)}(r,r\hx)/r}{(1+\hx\delta)^3} \,,
 }
where in the second line we integrated by parts once, and in the third line we defined $m = r \hat x$.

\subsection{Matrix model and volumes of five-cycles}

For any gauge invariant operator $X$, we should have \cite{Fabbri:1999hw} $R[X] = \pi \Vol(\Sigma_X)/ 6\Vol(Y)$, where by $\Sigma_X$ we denoted the 5-d submanifold of $Y$ defined by the equation $X=0$.  Using $\Vol(Y) = (\pi^4 / 24) \int d\hat x\, \hat \rho(\hat x)$, one can rewrite this equation as
 \es{VolX5Invar}{
  \Vol(\Sigma_X) = \frac{\pi^3}{4} \int d\hx\, \hat \rho(\hx) R[X] \,.
 }
For an operator $X$ that is not gauge invariant, such as a bifundamental field that transforms in $({\bf N}_a, \overline{\bf N}_b)$, $R[X]$ is not invariant under baryonic symmetries \eqref{Symmetries}, but $R[X] + \hat y_a(\hx) - \hat y_b(\hx)$ is.  So we suspect that
 \es{VolX5InvarAgain}{
   \Vol(\Sigma_X) = \frac{\pi^3}{4} \int d\hx\, \hat{\rho}(\hx) (R[X] + \hat y_a(\hx) - \hat y_b(\hx)) \,.
 }
We can think of this relation as a conjecture and prove the following result:  If $X$ is a chiral operator transforming in $({\bf N}_a, \overline{\bf N}_b)$, then for $\delta$ in a neighborhood of zero, let
  \es{VolX5Delta}{
\Vol_m(\Sigma_X,\delta) &= 
 \frac{\pi^3}{4} \int d\hx_\delta\, \hat{\rho}_\delta (\hx_\delta) (R[X] + \hat y_{a, \delta} (\hx_\delta) - \hat y_{b, \delta} (\hx_\delta)) \,. 
 }
 The following two statements are equivalent:
   \begin{enumerate}[I.]
     \item The conjecture~\eqref{resulttwo} holds for the given R-charges $R[X]$ and bare monopole charge $\Delta_m$. \label{FirstAssertionFive}
    \item  For any $\delta$ in a small enough neighborhood of zero, we have
     \es{VolXConjecture}{
      \Vol_m(\Sigma_X, \delta) = \Vol(\Sigma_X, \delta) \,,
     }
    where the volume $\Vol(\Sigma_X, \delta)$ is computed with the induced Sasakian metric on $Y$ that corresponds to the matter R-charges $R[X]$ and the bare monopole charge $\Delta_m + \delta$. \label{SecondAssertionFive}
    \end{enumerate}

The proof of this result is similar to that of the equivalence between \eqref{FirstAssertion} and \eqref{SecondAssertion} we discussed above, so we skip most of the details.  Using \eqref{NontrivialR}, one can check that
 \es{VolX5DeltaAgain}{
   \Vol_m(\Sigma_X,\delta) 
     = \frac{\pi^3}{4} \int d\hx\, \frac{\hat{\rho}(\hx) 
     (R[X] + \hat y_a(\hx) - \hat y_b(\hx))}{\left(1 + \hat x \delta \right)^3} \,. 
 }
 Defining $f_X(t, u)$ to be the Hilbert-Poincar\'e series for the ring of chiral operators obtained from the chiral ring by setting $X = 0$, and using the Martelli, Sparks, and Yau result \cite{Martelli:2006yb}
 \es{HilbertX}{
   \Vol(\Sigma_X,\delta) = \frac{\pi^3}{8} \lim_{t \to 1} (1-t)^3 f_{X}(t,t^{\delta}) \,,
 }
 one can show as in section~\ref{DeltaMGeometry} that
  \es{VolX5Hilbert}{
   \Vol(\Sigma_X,\delta) = \frac{\pi^3}{4} \int d\hx \frac{\lim_{r \to \infty} \psi_X^{(1,1)}(r,r\hx) / r}{(1+\hx\delta)^3} \,.
 }
Here, $\psi_X(r, m)$ denotes the number of chiral ring operators with $X = 0$, R-charge at most $r$, and monopole charge at most $m$, and can be approximated by a smooth function of homogeneous degree three.   By an argument analogous to the one in section~\ref{EQUIVRHO} it follows that the statements \eqref{FirstAssertionFive} and \eqref{SecondAssertionFive} are equivalent.

\subsection{A Consistency condition}

Note that gauge invariant operators in the quiver that have no monopole charge are constructed from closed paths of bifundamental fields ${\cal O}_\alpha$.  A consequence of our conjecture (\ref{results}b) is then that for a gauge invariant operator
$X = \prod_\alpha {\cal O}_\alpha$ with no monopole charge the following sum vanishes: 
\begin{equation}
\sum_{\alpha} \left[ \frac{\partial^2 \psi_{{\cal O}_\alpha}}{\partial r \partial m} - R[{\cal O}_\alpha] \frac{\partial^3 \psi}{\partial r^2 \partial m} \right ] = 0 \ .
\label{whatwewantinit}
\end{equation}
Of course $\sum_\alpha R[{\cal O}_\alpha] = R[X]$, and we can simplify this expression:
\begin{equation}
\sum_{\alpha} \frac{\partial^2 \psi_{{\cal O}_\alpha}}{\partial r \partial m} = R[X] \frac{\partial^3 \psi}{\partial r^2 \partial m} \ .
\label{whatwewant}
\end{equation}
We would like to show why (\ref{whatwewant}) must hold from geometric considerations 
alone.\footnote{%
 For a Calabi-Yau four fold, at large $r$ and $m$ we approximate
 $\psi(r,m)$ and $\psi_{{\cal O}_\alpha}(r,m)$ by
 homogeneous polynomials of degree four and three respectively. 
}

The number of gauge invariant operators of fixed R-charge $r$ and monopole charge $m$ that do not contain the operator $X$ is approximately
\begin{equation}
\psi^{(1,1)}_{X}(r,m) \approx \psi^{(1,1)}(r,m) - \psi^{(1,1)}(r-R[x], m)  \approx R[X] \psi^{(2,1)}(r,m) \ 
\label{firstcount}
\end{equation}
when $r \gg R[X]$ is large.

We can also use \eqref{HilbertX} to express the operator counts in
terms of volumes.  
For coordinates $x$ and $y$ on a compact space, the volume of the set of points $xy=0$ is the union of the set of points where $x=0$ with the set of points where $y=0$.
That the volumes are additive implies the identity 
\begin{equation}
\Vol(\Sigma_X, \delta) = \sum_{\alpha} \Vol (\Sigma_{\mathcal{O}_{\alpha}},\delta) \ .
\end{equation}
From this identity at large $r$ and the results in the earlier part of this section, we have
\begin{equation} \label{psiAdditive}
\psi_X^{(1,1)}(r, m) = \sum_{\alpha} \psi^{(1,1)}_{{\cal O}_\alpha}(r,m) \ .
\end{equation}

Combining \eqref{firstcount} with \eqref{psiAdditive}
yields \eqref{whatwewant}.

\section{Theories with non-chiral bifundamental fields}
\label{sec:centralexamples}

\subsection{${\cal N} = 2$ deformations of the necklace quivers and matrix model}
\label{NECKLACE}

\begin {figure} [!t]
  \centering
\newcommand {\svgwidth} {0.5\textwidth}

\begingroup
  \makeatletter
  \providecommand\color[2][]{%
    \errmessage{(Inkscape) Color is used for the text in Inkscape, but the package 'color.sty' is not loaded}
    \renewcommand\color[2][]{}%
  }
  \providecommand\transparent[1]{%
    \errmessage{(Inkscape) Transparency is used (non-zero) for the text in Inkscape, but the package 'transparent.sty' is not loaded}
    \renewcommand\transparent[1]{}%
  }
  \providecommand\rotatebox[2]{#2}
  \ifx\svgwidth\undefined
    \setlength{\unitlength}{356.7550293pt}
  \else
    \setlength{\unitlength}{\svgwidth}
  \fi
  \global\let\svgwidth\undefined
  \makeatother
  \begin{picture}(1,0.96074514)%
    \put(0,0){\includegraphics[width=\unitlength]{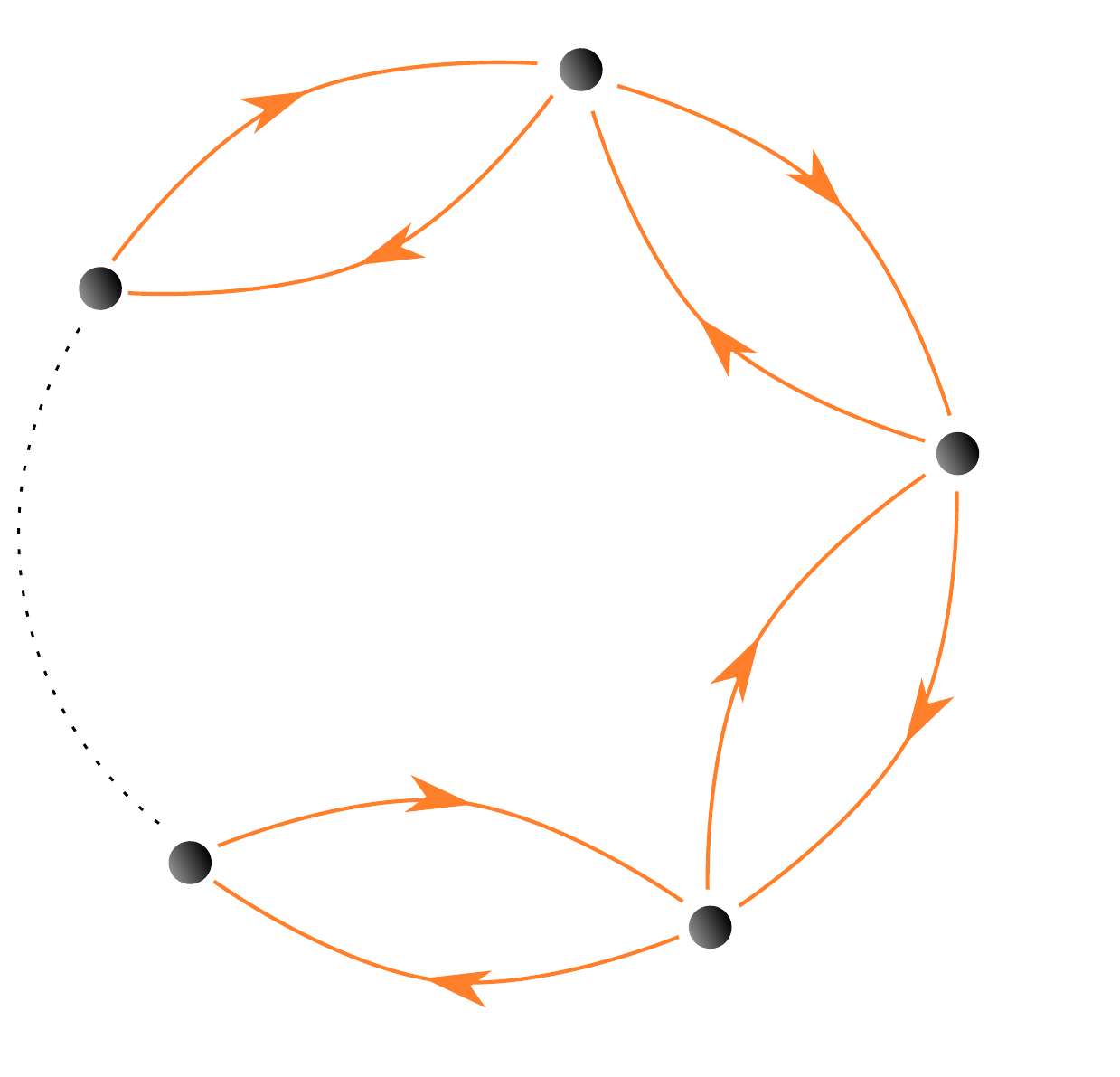}}%
    \put(0.52471435,0.97027939){\color[rgb]{0,0,0}\makebox(0,0)[lt]{\begin{minipage}{0.23385393\unitlength}\raggedright $k_d$\end{minipage}}}%
    \put(0.88264291,0.58508092){\color[rgb]{0,0,0}\makebox(0,0)[lt]{\begin{minipage}{0.13134264\unitlength}\raggedright $k_1$\end{minipage}}}%
    \put(0.64439419,0.11510091){\color[rgb]{0,0,0}\makebox(0,0)[lt]{\begin{minipage}{0.16658087\unitlength}\raggedright $k_2$\end{minipage}}}%
    \put(0.11770223,0.17208277){\color[rgb]{0,0,0}\makebox(0,0)[lt]{\begin{minipage}{0.17298791\unitlength}\raggedright $k_3$\end{minipage}}}%
    \put(-0.00333053,0.76851034){\color[rgb]{0,0,0}\makebox(0,0)[lt]{\begin{minipage}{0.19220874\unitlength}\raggedright $k_{d-1}$\\ \end{minipage}}}%
    \put(0.13901095,0.88727256){\color[rgb]{0,0,0}\makebox(0,0)[lb]{\smash{$A_d$}}}%
    \put(0.70602662,0.85844131){\color[rgb]{0,0,0}\makebox(0,0)[lb]{\smash{$A_1$}}}%
    \put(0.8341658,0.25939078){\color[rgb]{0,0,0}\makebox(0,0)[lb]{\smash{$A_2$}}}%
    \put(0.3216092,0.00951942){\color[rgb]{0,0,0}\makebox(0,0)[lb]{\smash{$A_3$}}}%
    \put(0.54905623,0.60856993){\color[rgb]{0,0,0}\makebox(0,0)[lb]{\smash{$B_1$}}}%
    \put(0.58429445,0.40354725){\color[rgb]{0,0,0}\makebox(0,0)[lb]{\smash{$B_2$}}}%
    \put(0.37286488,0.2882221){\color[rgb]{0,0,0}\makebox(0,0)[lb]{\smash{$B_3$}}}%
    \put(0.29277789,0.64380815){\color[rgb]{0,0,0}\makebox(0,0)[lb]{\smash{$B_d$}}}%
  \end{picture}%
\endgroup

  \caption {A necklace quiver gauge theory where the gauge sector consists of $d$ $U(N)$ gauge groups with Chern-Simons coefficients $k_a$ and the matter content consists of the bifundamental fields $A_a$ and $B_a$.
  \label{NecklaceFigure}}
\end {figure}
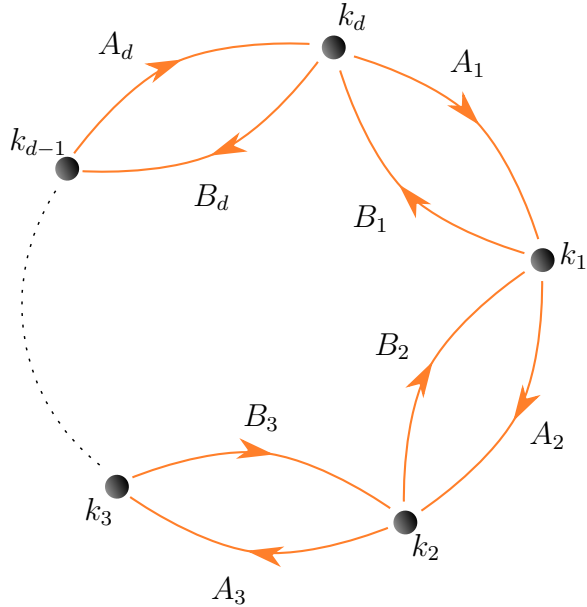

Our first field theory example consists of deformations of the necklace quiver gauge theories whose (undeformed) matrix models were also studied in \cite{Gulotta:2011si,Herzog:2010hf}. In ${\cal N} = 2$ notation, the field content of the necklace quiver theories consists of $d$ vector multiplets with Chern-Simons kinetic terms and coefficients $k_a$, and chiral multiplets $A_a$ and $B_a$ connecting the gauge groups into a necklace (see figure~\ref{NecklaceFigure}).  The superpotential
\be \label{NecklaceW}
W = \sum_{a=1}^d \frac{1}{k_a} \operatorname{tr}(B_{a+1} A_{a+1} - A_{a}B_{a})^2 
\ee
preserves ${\cal N} = 3$ supersymmetry.
For any given $k_a$ satisfying $\sum_{a = 1}^d k_a = 0$, the field theory is dual to $AdS_4 \times Y$ where $Y$ is a tri-Sasakian space, which is by definition the base of a hyperk\"ahler cone \cite{Jafferis:2008qz}.

While ${\cal N} = 3$ SUSY restricts the R-charges of $A_a$ and $B_a$ to be $1/2$, in this section we examine what happens if we make more general R-charge assignments for the $A_a$ and $B_a$ fields that break ${\cal N} = 3$ down to ${\cal N} = 2$.  These R-charge assignments are required to preserve the marginality of the superpotential \eqref{NecklaceW}.  This condition implies that for generic values of the CS levels $k_a$, namely if there are no cancellations between the various terms in \eqref{NecklaceW}, we must have $R[B_a] = 1 - R[A_a]$.  The matrix model free energy functional is in this case
 \es{ArbitraryRModel}{
  &\tilde F[\rho, y_a, \mu] = 2 \pi N^{3/2} \int dx\, \rho x \sum_{a=1}^d q_a \delta y_a 
   + 2 \pi \Delta_m N^{3/2} \int dx\, \rho x \\
   &\quad{}- \pi N^{3/2} \int dx\, \rho^2 \sum_{a=1}^d \left(\delta y_a -R[A_a] \right) \left(\delta y_a + R[B_a] \right) 
   - 2 \pi N^{3/2} \mu \left(\int dx\, \rho - 1 \right) \,,
 }
where $\delta y_a = y_{a-1} - y_a$, $k_a = q_{a+1} - q_a$.  As per the discussion after eq.~\eqref{FFunctional}, the equations of motion for $\delta y_a$ following from \eqref{ArbitraryRModel} hold only when $-R[B_a] < \delta y_a < R[A_a]$.  It is possible to have $\delta y_a = R[A_a]$ or $\delta y_a = - R[B_a]$, but in that case we should not impose the equation of motion for that particular $\delta y_a$.

Using the rescalings \eqref{hatDefs}, one can write the solution of the equations of motion following from \eqref{ArbitraryRModel} as
 \es{SolnR}{
  \hat \rho(\hat x) &=  s_L(\hat x) - s_S(\hat x)  \,, \quad 
   \delta \hat y_a(\hat x) = \frac{R[A_a] - R[B_a]}{2} + 
     \frac{1}{2} \frac{\abs{s_L(\hat x) + q_a \hat x} -  \abs{s_S(\hat x)+ q_a \hat x}}{s_L(\hat x) - s_S(\hat x)} \,,
 }
where $s_L(\hat x) \geq s_S(\hat x)$ are the two solutions of the equation
 \es{sEq}{
  s(\hat x) c_1 + \hat x c_2
     +  \sum_{a = 1}^d \abs{ s(\hat x) + \hat x q_a}= 2 \,,
 }
with
 \es{DeltaDefs}{
   c_1 \equiv \sum_{a=1}^d \left( R[A_a] - R[B_a] \right) \,, \qquad
  c_2 \equiv  2 \Delta_m + \sum_{a=1}^d q_a \left( R[A_a] - R[B_a]\right)
    \,.
 }
The constraint imposed by varying $\tilde F$ with respect to $\mu$ is $\int d\hat x\, \hat \rho(\hat x) = 1/\mu^2$.

We have encountered a solution of this type in \cite{Gulotta:2011si} in the case where $R[A_a] = R[B_a] = 1/2$ and $\Delta_m =0$. As in \cite{Gulotta:2011si}, one can think of eq.~\eqref{sEq} as defining the boundary of a polygon
 \es{PDef}{
   {\cal P} = \left\{ (\hat x, s)\in \R^2 :  
   s c_1 + \hat x c_2
     +  \sum_{a = 1}^d \abs{ s + \hat x q_a} \leq 2
      \right\} \,.
 }
The quantity $\hat \rho(\hat x) =  s_L(\hat x) - s_S(\hat x)$ can then be interpreted as the thickness of a constant $\hat x$ slice ${\cal P}_{\hat x}$ through this polygon, $\hat \rho(\hat x) = \Length({\cal P}_{\hat x})$.  Consequently, $ \int d\hat x\, \hat \rho(\hat x) = \Area({\cal P})$ and
 \es{VolmNecklace}{
  \Vol_m(Y) = \frac{\pi^4}{24 \mu^2} = \frac{\pi^4}{24} \Area({\cal P}) \,.
 }
See appendix~\ref{FMAXIMIZATIONNECKLACE} for a proof that the ${\cal N} = 3$ R-charge assignments minimize $\Vol_m(Y)$ or, equivalently, maximize $F$.  Just like $\hat \rho(\hat x)$, the quantities $\hat \rho(\hat x) \delta \hat y_a (\hat x)$ can also be given geometrical interpretations:
 \es{rhodeltayInterpretation}{
   \hat \rho(\hat x) (\delta \hat y_a(\hat x)  + R[B_a] ) &=  \Length \left( {\cal P}_{\hat x} \cap \{s + q_a \hat x \geq 0 \} \right) \,, \\
  \hat \rho(\hat x) (-\delta \hat y_a(\hat x)  + R[A_a] ) &= \Length \left( {\cal P}_{\hat x} \cap \{s + q_a \hat x \leq 0 \}\right) \,.
 }

The equations above were written in a way that makes manifest the invariance under the flat directions exhibited in \eqref{Symmetries}.  Indeed, while in writing the free energy functional \eqref{ArbitraryRModel} we assumed $R[A_a]$ and $\Delta_m$ to be independent, we see that the eigenvalue density $\hat \rho(\hat x)$ and the quantities appearing on the LHS of \eqref{rhodeltayInterpretation} depend non-trivially only on the linear combinations $c_1$ and $c_2$ that were defined in \eqref{DeltaDefs}.   These are the only linear combinations of $R[A_a]$ and $\Delta_m$ that are invariant under all symmetries in \eqref{Symmetries}.  The reason why we were able to find two such linear combinations at all is that the spaces $Y$ have generically two $U(1)$ isometries that commute with $U(1)_R$.

\subsection{Operator counting for necklace quivers}

We now relate the matrix model quantities $\hat \rho(\hat x)$ and $\hat \rho(\hat x) \delta \hat y_a$ from the previous section to numbers of operators in the chiral ring of the gauge theory when $N=1$.  In \cite{Gulotta:2011si} we provided such a relation in the case $R[A_a] = R[B_a] = 1/2$ and $\Delta_m = 0$, and the argument presented in that paper holds, with minor modifications, for the more general R-charge assignments considered in this paper.  As explained in~\cite{Gulotta:2011si}, gauge invariant operators can be constructed out of the bifundamental fields $A_a$ and $B_a$ and the diagonal monopole operators $T^{(m)}$, and they are
 \es{ODef}{
   \mathcal{O}(m,s,i,j)  &= T^{(m)} C_1^{m q_1 + s} C_2^{m q_2 + s} \cdots C_d^{m q_d + s} (A_1 B_1)^i (A_2 B_2)^j \,, \\
    C_a^{m q_a + s} &\equiv
    \begin{cases}
      A_a^{mq_a + s} \  & \mbox{if  } mq_a +s > 0 \  \\
      B_a^{-m q_a - s} \ & \mbox{if  } mq_a + s < 0 \ 
    \end{cases} \,.
 } 
The labels $m$ and $s$ run over all integers, while $i$ and $j$ should be nonnegative integers.

Let $\psi(r, m)$ ($\psi_0(r, m)$) be the number of operators ${\cal O}(m, s, i, j)$ (${\cal O}(m, s, 0, 0)$) with R-charge at most $r$ and monopole charge at most $m$.  In \cite{Gulotta:2011si} we showed that at large $r$ and $m$ we have $\psi^{(2, 0)} (r, m) \approx \psi_0(r, m)$.  This relation holds for the more general R-charge assignments too because the only assumption needed to prove it was $R[A_1 B_1] = R[A_2 B_2] = 1$, which we still assume. A simple computation yields
 \es{RO}{
  R[{\cal O}(m, s, 0, 0)] &= m \Delta_m + \sum_k R[C_k] \abs{m q_a + s}
    = \frac{1}{2} \left[ s c_1 + m c_2  +  \sum_{a = 1}^d \abs{ s + m q_a} \right] \,,
 }
where $c_1$ and $c_2$ are as defined in \eqref{DeltaDefs}.  Using this formula one can check, as in \cite{Gulotta:2011si}, that $\psi^{(2,1)}(r, r \hat x) / r \approx \psi_0^{(0,1)}(r, r \hat x) / r$ is indeed given by the length of the slice ${\cal P}_{\hat x}$ through ${\cal P}$.  We have therefore verified explicitly eq.~\eqref{resultone} for the necklace quivers at non-critical R-charges.

Let $\psi_{X_a}(r, m)$ be the number of chiral operators with R-charge at most $r$ and monopole charge at most $m$ that are nonzero when $X_a = 0$.  As in \cite{Gulotta:2011si}, we have that $\psi_{X_a}^{(1, 0)}(r, m)$ equals the number of operators of the form ${\cal O}(m, s, 0, 0)$ with R-charge at most $r$ and monopole charge at most $m$ with the extra constraint that $m q_a + s \leq 0$ if $X_a = A_a$ and $m q_a + s \geq 0$ if $X_a = B_a$.  As argued in \cite{Gulotta:2011si}, these extra constraints imply that when $r$ is large $\psi_{X_a}^{(1, 1)}(r, r \hat x)/r$ is given by the length of the intersection between the slice ${\cal P}_{\hat x}$ and the half-plane $s + q_a \hat x \geq 0$ if $X_a = B_a$ or $s + q_a \hat x \leq 0$ if $X_a = A_a$.  Comparing with eq.~\eqref{rhodeltayInterpretation} we see that the necklace quivers at arbitrary R-charges also obey our second conjecture \eqref{resulttwo}.

\subsection{Flavored necklace quivers}

The discussion in the previous two subsections can be generalized by including flavor fields that interact with the existing matter fields through the superpotential
 \es{WFlavoredNecklace}{
  \delta W \sim \sum_{a=1}^d \tr \left[  \sum_{j=1}^{n_{a}} \tilde q_j^{(a)} A_a  q_j^{(a)} 
   + \sum_{j=1}^{m_{a}} Q_j^{(a)} B_a \tilde Q_j^{(a)} \right] \,.
 }
Given that the $A_a$ transform in $(\overline{\bf N}_{a-1}, {\bf N}_{a})$ and the $B_a$ transform in the conjugate representation $({\bf N}_{a-1}, \overline{\bf N}_{a})$, for eq.~\eqref{WFlavoredNecklace} to make sense we must take $q_j^{(a)}$, $\tilde q_j^{(a)}$, $Q_j^{(a)}$, and $\tilde Q_j^{(a)}$ to transform in ${\bf N}_{a-1}$, $\overline{\bf N}_{a}$, $\overline{\bf N}_{a-1}$, and ${\bf N}_{a}$, respectively.

We discussed a superpotential of this form at the end of section~\ref{NonCriticalR}, where we found that the effect of including the flavor fields was that the CS levels $k_a$ and $\Delta_m$ of the unflavored model were replaced by $(\sgn x) g_a[T^{(\sgn x)}]$ and $(\sgn x) R[T^{(\sgn x)}]$, respectively.  Eqs.~\eqref{TGauge} and~\eqref{TRcharge} applied to our flavored necklace quivers give
 \es{gRNecklace}{
  k_a &\to (\sgn x) g_a[T^{(\sgn x)}] = k_a + \frac{\sgn x}{2} \left(n_a - m_a - n_{a+1} + m_{a+1} \right)  \,, \\
  \Delta_m &\to (\sgn x) R[T^{(\sgn x)}] = \Delta_m + \frac{\sgn x}{2} \sum_a \left(n_a R[A_a] + m_a R[B_a]  \right) \,.
 }
From $k_a = q_{a+1} - q_a$ we further have
 \es{qNecklace}{
  q_a \to q_a - \frac{\sgn x}{2} \left(n_a - m_a \right) \,.
 }
We believe that all the formulas presented in the previous two subsections continue to hold for the flavored theory if one makes the above three replacements.  In particular, the relation between the matrix model quantities and operator counting we conjectured in eq.~\eqref{results} continues to hold, and the volume of the 7-d space $Y$ is still proportional to the area of a polygon ${\cal P}$ of the type \eqref{PDef}.

\subsection{Flavored $\mathcal{N}=8$ theory and its matrix model}
\label{sec:flavoredN8}

We broaden our scope of examples and verify (\ref{results}) for maximally supersymmetric Yang-Mills theory to which we add flavor.  The theory has one gauge group and three adjoint fields $X_i$, $i = 1, 2, 3$ coupled to $n_1 + n_2 + n_3$ pairs of fundamental fields through the superpotential 
\es{FlavoredSuperpot}{
  W \sim  \tr \left[X_1[X_2, X_3] +  \sum_{j=1}^{n_1} q_j^{(1)} X_1 \tilde q_j^{(1)}
    + \sum_{j=1}^{n_2} q_j^{(2)} X_2 \tilde q_j^{(2)}
    + \sum_{j=1}^{n_3} q_j^{(3)} X_3 \tilde q_j^{(3)}\right] \,.
 }
 The corresponding matrix model was solved in  \cite{Jafferis:2011zi} in the large $N$ limit.  We review their solution for $\rho(x)$.  In the next subsection, we will compare $\rho(x)$ with the distribution of operators in the chiral ring and show that (\ref{resultone}) holds.  In this case, eq.~(\ref{TTtildeOPE}) takes the form $ T^{(1)}  T^{(-1)} = X_1^{n_1} X_2^{n_2} X_3^{n_3} $ \cite{Jafferis:2009th, Benini:2009qs}.  To keep the notation concise, we define $\Delta_i \equiv R[X_i]$, $\Delta \equiv R[T^{(1)}]$, and $\tilde \Delta \equiv R[T^{(-1)}]$.  The matrix model free energy functional is then
 \es{FreeOneNode}{
  \tilde F[\rho] = \pi N^{3/2} \left[ \int dx \rho \left( \Delta_1 \Delta_2 \Delta_3   \rho
    + ( \Delta + \tilde \Delta )  \abs{x} 
    +  (\Delta-\tilde \Delta)  x \right) - 2  \mu \left( \int dx \, \rho - 1 \right) \right]
    \,.
 }
As before, we define the hatted quantities (\ref{hatDefs}).  The eigenvalue density $\hr(\hx)$ is
  \es{rhoOne}{
   \hr(\hx) = 
    \begin{cases}
     \frac{1 - \hx \Delta} {\Delta_1 \Delta_2 \Delta_3} & \text{if $0< \hx < \frac{1}{\Delta}$} \,, \\
     \frac{1 + \hx \tilde \Delta} {\Delta_1 \Delta_2 \Delta_3} & \text{if $-\frac{1}{\tilde \Delta} < \hx<0$} \,, \\
     0 & \text{otherwise} \,,
    \end{cases}
 }
 which agrees with (4.8) of \cite{Jafferis:2011zi}.

\subsection{Operator counting in flavored $\mathcal{N}=8$ theory}

The gauge-invariant operators built out of diagonal monopole operators and adjoint fields in this theory are $\tr [ T^{(m)} X_1^{a_1} X_2^{a_2} X_3^{a_3} ]$.  The R-charges of these operators are
 \es{RchargesOne}{
  R[T^{(m)} X_1^{a_1} X_2^{a_2} X_3^{a_3}] &= \begin{cases}
   m \Delta+ \sum_{i=1}^3 a_i \Delta_i  & m \geq 0 \,,
   \\
   -m \tilde \Delta + \sum_{i=1}^3 a_i \Delta_i & m < 0 \,.
   \end{cases} 
 }

 Let $\psi(r, m)$ be the number of operators with R-charge smaller than $r$ and monopole charge smaller than $m$.  To match with $\rho(x)$, we want to calculate $\partial^3 \psi / \partial r^2 \partial m$ at large $r$.  It is easiest to start by calculating the derivative $\partial \psi / \partial m$ which equals the number of operators with R-charge smaller than $r$ and monopole charge equal to  $m$.   For $m>0$, at large $r$ the number of operators $\tr [ T^{(m)} X_1^{a_1} X_2^{a_2} X_3^{a_3} ]$ is approximately equal to the volume of a tetrahedron with sides of length $(r - m \Delta)/\Delta_i$;  similarly, for $m<0$, the number of operators  is equal to the volume of a tetrahedron with sides of length $(r + m \tilde \Delta)/\Delta_i$.  We thus have
  \es{Gotpsi}{
   \frac{\partial \psi}{\partial m}
    = \begin{cases}
     \frac{(r - m \Delta)^3} {6 \Delta_1 \Delta_2 \Delta_3} & \text{if $0< m < \frac{r}{\Delta}$} \,, \\
     \frac{(r + m \tilde \Delta)^3} {6 \Delta_1 \Delta_2 \Delta_3} & \text{if $-\frac{r}{\tilde \Delta} < m<0$} \,, \\
     0 & \text{otherwise} \,.
    \end{cases}
  }
  Taking two derivatives with respect to $r$, we find agreement with \eqref{rhoOne} and confirmation of the conjecture (\ref{resultone}).

\subsection{Other examples}

We presented flavored ${\mathcal N}=8$ in the main text  because of its simplicity. 
 One disadvantage of this example is that it possesses a single $U(N)$ factor and so we could not compute a $\delta y$ and check (\ref{results}b).  To remedy this problem, in appendix \ref{app:furtherexamples} we consider two more complicated examples.  The first of these is ABJM Chern-Simons theory (a theory with two gauge groups) \cite{Aharony:2008ug} to which we add flavor.
The second example has four gauge groups (see figure \ref{C3Z2Z2Figure}).  When a four-dimensional gauge theory has the field content of this second example, the Abelian moduli space is a $\mathbb{Z}_2 \times \mathbb{Z}_2$ orbifold of $\mathbb{C}^3$.  Thus, with some abuse of notation, we refer to this second example as the $\mathbb{Z}_2 \times \mathbb{Z}_2$ orbifold theory.

The verification of (\ref{results})
requires on the one hand calculating $\rho(x)$ and $\delta y(x)$ using the large $N$ limit of the matrix model (\ref{FFunctional}) and on the other counting operators in the chiral ring.  
We have two methods at our disposal for this counting.
One may count the operators directly as we did above.  Because the moduli space is toric for these last three examples, 
the direct approach has some generic features which we review in appendix \ref{app:toric}.
In section \ref{sec:proofs}, we  presented an indirect counting method that involved 
calculating $\Vol(Y, \delta)$ (\ref{VolY7Hilbert}) and $\Vol(\Sigma_X, \delta)$ (\ref{VolX5Hilbert}) as a function of $\rho(x)$ and $\delta y( x)$.

\section{Theories with chiral bifundamental fields}
\label{sec:c3z3}

\subsection{Noncancellation of long-range forces}

As noted in \cite{Jafferis:2011zi}, 
the functional \eqref{FFunctional} does not appear to describe the large $N$ limit of gauge theories with chiral bifundamental fields.  To derive (\ref{FFunctional}), it was assumed that the long-range forces on the eigenvalues cancel.  But for theories with chiral bifundamentals, there is no such cancellation.

The long-range forces at issue come from the interactions between the eigenvalues, both within a vector multiplet and between vector multiplets connected by a bifundamental field $X_{ab}$
\cite{Jafferis:2011zi}:
\es{Fwithin}{
F_{i,\mathrm{self}}^{(a)} &= \sum_{j \ne i} \coth \pi (\lambda_i^{(a)}-\lambda_j^{(a)}) \ , \\
F_{i,\mathrm{inter}}^{(a,b)} &= \sum_j \left[ \frac{R[X_{ab}]-1-i(\lambda_i^{(b)}-\lambda_j^{(a)})}{2} \right] \coth \pi \left(\lambda_i^{(b)}-\lambda_j^{(a)}-i(1-R[X_{ab}])\right)  \ , \\
F_{i,\mathrm{inter}}^{(b,a)} &= \sum_j \left[ \frac{R[X_{ba}]-1+i(\lambda_i^{(b)}-\lambda_j^{(a)})}{2} \right] \coth \pi \left(\lambda_i^{(b)}-\lambda_j^{(a)}+i(1-R[X_{ba}]) \right) \ .
}
If $\abs{\lambda_i^{(a)}-\lambda_j^{(b)}} \gg 1$, then we may approximate
$\coth x \approx \sgn \Re x$.
The long-range forces are the forces \eqref{Fwithin}
with $\coth$ replaced by $\sgn \Re$.
For theories with non-chiral bifundamentals and equal ranks,  the long-range forces cancel out
when
$\Re \lambda_i^{(a)} = \Re \lambda_i^{(b)}$ for all $i,a,b$ and 
(\ref{BetaCond}) is satisfied.  In general, the long-range forces on $\lambda_i^{(a)}$ cancel out only when
\begin{equation}
\sum_{b} (R[X_{ab}]-1 + 
y_{b,j}
) + \sum_{b} (R[X_{ba}]-1 - 
y_{b,j}
) = -2 \ . 
\label{RChargeChiral}
\end{equation}
Thus the free energy functional \eqref{FFunctional} is correct for theories with chiral bifundamentals only if the $y_a(x)$ satisfy some constraints.

\subsection{Operator counting for the $\mathbb{C}^3 / \mathbb{Z}_3$ theory}
\label{app:c3z3counting}

\begin {figure} [!t]
  \centering
\newcommand {\svgwidth} {0.37\textwidth}

\begingroup
  \makeatletter
  \providecommand\color[2][]{%
    \errmessage{(Inkscape) Color is used for the text in Inkscape, but the package 'color.sty' is not loaded}
    \renewcommand\color[2][]{}%
  }
  \providecommand\transparent[1]{%
    \errmessage{(Inkscape) Transparency is used (non-zero) for the text in Inkscape, but the package 'transparent.sty' is not loaded}
    \renewcommand\transparent[1]{}%
  }
  \providecommand\rotatebox[2]{#2}
  \ifx\svgwidth\undefined
    \setlength{\unitlength}{190.82110596pt}
  \else
    \setlength{\unitlength}{\svgwidth}
  \fi
  \global\let\svgwidth\undefined
  \makeatother
  \begin{picture}(1,0.87485409)%
    \put(0,0){\includegraphics[width=\unitlength]{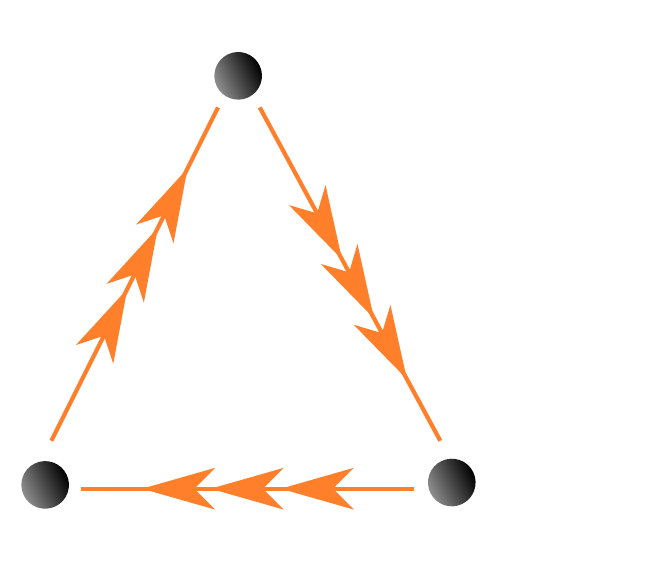}}%
    \put(0.68548332,0.0961778){\color[rgb]{0,0,0}\makebox(0,0)[lt]{\begin{minipage}{0.32341447\unitlength}\raggedright $k_2$\end{minipage}}}%
    \put(0.32907249,0.89275393){\color[rgb]{0,0,0}\makebox(0,0)[lt]{\begin{minipage}{0.24555537\unitlength}\raggedright $k_1$\end{minipage}}}%
    \put(-0.00626405,0.09301299){\color[rgb]{0,0,0}\makebox(0,0)[lt]{\begin{minipage}{0.32341447\unitlength}\raggedright $k_3$\end{minipage}}}%
    \put(0.05662213,0.46116494){\color[rgb]{0,0,0}\makebox(0,0)[lb]{\smash{$A_{31, i}$}}}%
    \put(0.26624255,0.0419241){\color[rgb]{0,0,0}\makebox(0,0)[lb]{\smash{$A_{23, i}$}}}%
    \put(0.55971113,0.46116494){\color[rgb]{0,0,0}\makebox(0,0)[lb]{\smash{$A_{12, i}$}}}%
  \end{picture}%
\endgroup

  \caption {The quiver for the $\C^3/\Z_3$ theory.  When the CS levels are $(2k, -k, -k)$ this field theory is believed to be dual to $AdS_4 \times M^{1, 1,1} / \Z_k$.
  \label{C3Z3Figure}}
\end {figure}
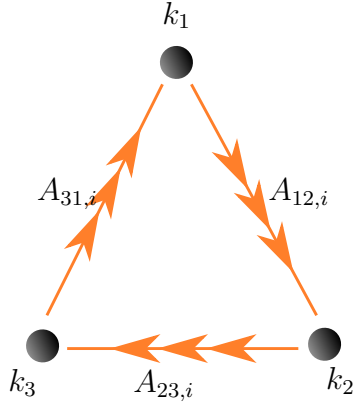

To investigate what the matrix model for a chiral theory should give in the large $N$ limit, 
we study the $U(N)^3$ Chern-Simons theory described by the quiver
in figure \ref{C3Z3Figure}.  Let the Chern-Simons coefficients be $(k_1, k_2, k_3)$ such that 
$k_1 + k_2 + k_3 = 0$.  
We will assume $k_1>0$, $k_2<0$, $k_3<0$.
The moduli space is a K\"ahler quotient of $\C^5$ with weights
$(\frac{1}{3}(k_++k_-), \frac{1}{3}(k_++k_-), \frac{1}{3}(k_++k_-),-k_+,-k_-)$, where we define $k_- = k_1 - k_2$ and $k_+ = k_1 - k_3$.

There is a superpotential of the form
\begin{equation}
W \sim \tr \left[ \epsilon_{ijk} A_{31,k} A_{23,j} A_{12,i}   \right]\ ,
\end{equation}
and a monopole relation $T^{(1)} T^{(-1)} = 1$.  
We let
$R[A_{ij,1}]=\Delta_x$, $R[A_{ij,2}]=\Delta_y$, $R[A_{ij,3}]=\Delta_z$, with
$\Delta_x + \Delta_y + \Delta_z = 2$ as
any other choice of R-charges may be transformed into this choice by a
transformation of the form \eqref{Symmetries}.
We denote $R[T^{(1)}] = - R[T^{(-1)}] = \Delta$.

The gauge invariant operators have the form 
 \es{OpsC3Z3}{
T^{(m)}  \prod_{i=1}^3 \prod_{j=1}^3 (A_{i(i+1), j})^{n_{i(i+1),j}}   \,.
 }
To be gauge invariant, for $m \geq 0$ we must impose
 $\sum_j n_{12,j} = m k_1+s$, $\sum_j n_{23,j} = m k_1 + m k_2+s$ and $\sum_j n_{31,j} = s$
 and for $m<0$ we must impose $\sum_j n_{23,j} = m k_2+s$, $\sum_j n_{31,j} = m k_2 + m k_3+s$, and 
$ \sum_j n_{12,j} = s$.  
Given the R-charge assignments, it is convenient to introduce 
$n_j = \sum_i n_{i(i+1),j}$. 
Each gauge invariant operator corresponds to a quadruple $(n_1, n_2, n_3, m)$ such that
$\sum_j n_j = m k_{\sgn(m)} + 3s$ and $m$ is bounded between $-\sum_j n_j / k_-$ and
$\sum_j n_j / k_+$.

Given the description of the gauge invariant operators, it is now a straightforward task to count them by either the direct method described in appendix \ref{app:toric} or the indirect method described in section \ref{sec:proofs}.  For $\Delta_x \ge \Delta_y \ge \Delta_z$
a piecewise expression for $\hr(\hx)$ is:
\begin{equation}
\label{rhoc3z3}
\hat{\rho}(\hx) = \left\{ \begin{array}{ll}
 0\,, 
   & \hx \le - \frac{1}{k_-\Delta_z-\Delta} \,, \\
 \frac{1+(k_- \Delta_z - \Delta) \hx}{3\Delta_z(\Delta_x-\Delta_z)(\Delta_y-\Delta_z)}\,,
   & -\frac{1}{k_-\Delta_z-\Delta} \le \hx \le -\frac{1}{k_-\Delta_y-\Delta} \,, \\
 \frac{(\Delta_x - \Delta_y - \Delta_z)(1-\Delta \hx) - \Delta_y \Delta_z k_-\hx}{3(\Delta_x-\Delta_y)(\Delta_x-\Delta_z) 
     \Delta_y \Delta_z} \,,
   & -\frac{1}{k_-\Delta_y-\Delta} \le \hx \le -\frac{1}{k_-\Delta_x-\Delta} \,, \\
 \frac{1-\Delta \hx}{3 \Delta_x \Delta_y \Delta_z}\,,
   & -\frac{1}{k_-\Delta_x-\Delta} \le \hx \le \frac{1}{k_+\Delta_x+\Delta} \,, \\
 \frac{(\Delta_x - \Delta_y - \Delta_z)(1-\Delta \hx) + \Delta_y \Delta_z k_+\hx}{3(\Delta_x-\Delta_y)(\Delta_x-\Delta_z) 
     \Delta_y \Delta_z} \,,& \frac{1}{k_+\Delta_x+\Delta} \le \hx \le \frac{1}{k_+\Delta_y+\Delta} \,, \\
 \frac{1-(k_+\Delta_z+\Delta) \hx}{3\Delta_z(\Delta_x-\Delta_z)(\Delta_y-\Delta_z)} \,, 
   & \frac{1}{k_+\Delta_y+\Delta} \le \hx \le \frac{1}{k_+\Delta_z+\Delta} \,, \\
 0 \,, 
   & \frac{1}{k_+\Delta_z+\Delta} \le \hx \,. \\
\end{array} \right.
\end{equation}
We note three odd things about (\ref{rhoc3z3}):
1)
If $\Delta_x = \Delta_y = \Delta_z$, $\hat{\rho}$ has a delta function
at $-\frac{1}{k_- \Delta_x - \Delta}$ and $\frac{1}{k_+ \Delta_x + \Delta}$.
2) In contrast to nonchiral examples, $\hat \rho(\hat x)$ while still piecewise linear is no longer a convex function of $\hat x$.  
3)
The matrix model (\ref{FFunctional}) gives the same 
result for $\hat \rho$ in the central region despite the fact that the long range forces do not cancel.
(In other regions and for $\delta \hat  y_{ab}$, the matrix model results are different.)

Now we set $A_{23,1}$ to zero.  The nonzero operators are those with no $A_{i(i+1),1}$ 
fields.  As a piecewise function:
\begin{equation}
\label{y3y2diff}
\hat y_3(\hx) - \hat y_2(\hx) = \left\{ \begin{array}{ll}
  -\Delta_z\,,
    & -\frac{1}{k_-\Delta_z-\Delta} \le \hx \le -\frac{1}{k_-\Delta_y-\Delta} \,, \\
  \frac{\Delta_y \Delta_z (1+(k_-\Delta_x -\Delta) \hx)}
    {(\Delta_x - \Delta_y - \Delta_z)(1-\Delta \hx) -  \Delta_y \Delta_z k_- \hx} \,,
    & -\frac{1}{k_-\Delta_y-\Delta} \le \hx \le -\frac{1}{k_-\Delta_x-\Delta} \,, \\
  0\,,
    & -\frac{1}{k_-\Delta_x-\Delta} \le \hx \le \frac{1}{k_+\Delta_x+\Delta} \,, \\
  \frac{\Delta_y \Delta_z (1-(k_+\Delta_x+\Delta) \hx)}
    {(\Delta_x - \Delta_y - \Delta_z)(1-\Delta \hx) + \Delta_y \Delta_z k_+ \hx} \,,
    & \frac{1}{k_+\Delta_x+\Delta} \le \hx \le \frac{1}{k_+\Delta_y+\Delta} \,, \\
  -\Delta_z \,,
    & \frac{1}{k_+\Delta_y+\Delta} \le \hx \le \frac{1}{k_+\Delta_z+\Delta} \,. \\
\end{array} \right.
\end{equation}

Finally, we set $A_{31,1}$ to zero.  The nonzero operators are those with no $A_{i(i+1),1}$'s, and those
with $m \ge 0, n_x+n_y+n_z=k_+ m$.  As a piecewise function:
\begin{equation}
\label{y2y1diff}
  \hat y_1(\hx) - \hat y_3(\hx) = \left\{ \begin{array}{ll}
   -\Delta_z \,,
      & -\frac{1}{k_-\Delta_z-\Delta} \le \hx \le -\frac{1}{k_-\Delta_y-\Delta} \,, \\
   \frac{\Delta_y \Delta_z (1+(k_-\Delta_x-\Delta) \hx)}
     {(\Delta_x - \Delta_y - \Delta_z)(1-\Delta \hx) -  \Delta_y \Delta_z k_- \hx} \,,
     & -\frac{1}{k_-\Delta_y-\Delta} \le \hx \le -\frac{1}{k_-\Delta_x-\Delta} \,, \\
   0 \,,
     & -\frac{1}{k_-\Delta_x-\Delta} \le \hx \le \frac{1}{k_+\Delta_x+\Delta} \,, \\
   -2\frac{\Delta_y \Delta_z (1-(k_+\Delta_x +\Delta) \hx)}
     {(\Delta_x - \Delta_y - \Delta_z) (1-\Delta \hx) +  \Delta_y \Delta_z k_+ \hx}\,, 
     & \frac{1}{k_+\Delta_x + \Delta} \le \hx \le \frac{1}{k_+\Delta_y+\Delta} \,, \\
   2\Delta_z\,, 
     & \frac{1}{k_+\Delta_y+\Delta} \le \hx \le \frac{1}{k_+\Delta_z+\Delta}\,. \\
\end{array} \right.
\end{equation}
The result for $\hat y_2(\hat x) - \hat y_1(\hat x)$ follows by taking the difference of (\ref{y3y2diff}) and (\ref{y2y1diff}).  We have checked that the operator counts where we set each of the remaining seven bifundamental fields to zero in turn yield the same results for the differences in the $\hat y$'s.

\subsection{Particular case:  the cone over $M^{1, 1, 1}/\Z_k$}

Consider the case where the internal space $Y$ is $M^{1, 1, 1} / \Z_k$.  It was proposed in \cite{Martelli:2008si, Hanany:2008cd, Franco:2009sp} that the dual field theory is the one in figure~\ref{C3Z3Figure} with CS levels $k_1 = 2 k$ and $k_2 = k_3 = -k$, so $k_+ = k_- = 3 k$.  As a function of the trial R-charges, the volume of $Y$ is
 \es{VolYM111}{
  \Vol(Y) = \frac{\pi^4}{24} \int d\hat x\, \hat \rho(\hat x)
   = \frac{3 k^3 \pi^4 \left(\Delta^2 + 9 k^2 (\Delta_x \Delta_y + \Delta_x \Delta_z + \Delta_y \Delta_z\right)}
    {8 \left(9 k^2 \Delta_x^2 - \Delta^2\right) \left(9 k^2 \Delta_y^2 - \Delta^2\right) \left(9 k^2 \Delta_z^2 - \Delta^2\right)} \,.
 }
Under the constraint $\Delta_x + \Delta_y + \Delta_z = 2$, this expression is maximized for $\Delta_x = \Delta_y = \Delta_z = 2/3$ and $\Delta = 0$, and the maximum is $9 \pi^4 / (128 k)$, which is the volume of $M^{1, 1, 1} / \Z_k$ \cite{Fabbri:1999hw}.

For the critical R-charges, our predicted eigenvalue density is
 \es{rhoM111Predicted}{
  \hat \rho(\hat x) &=   \frac{9}{8} \theta \left(\frac{1}{2k} - \abs{\hat x} \right) +  \frac{9}{32k} \delta\left(\hat x + \frac{1}{2k} \right)
   + \frac{9}{32k} \delta\left(\hat x - \frac{1}{2k} \right) \,, \\
  \hat \rho(\hat x) \left( \hat y_3(\hat x) - \hat y_2(\hat x) \right) &= -\frac{3}{16 k} \delta\left(\hat x + \frac{1}{2k} \right)
   -\frac{3}{16 k} \delta\left(\hat x - \frac{1}{2k} \right) \,, \\
  \hat \rho(\hat x) \left( \hat y_2(\hat x) - \hat y_1(\hat x) \right) &= -\frac{3}{16 k} \delta\left(\hat x + \frac{1}{2k} \right)
   +\frac{3}{8 k} \delta\left(\hat x - \frac{1}{2k} \right) \,.
 }
The volumes of the five-cycles corresponding to the bifundamental fields are
 \es{Vols}{
  \Vol(\Sigma_{A_{23, a}} ) = \frac{\pi^3}{4} \int dx\, \hat \rho(\hat x) \left( \hat y_3(\hat x) - \hat y_2(\hat x)  + \frac{2}{3}\right)
   = \frac{3 \pi^3}{16 k} \,, \\
  \Vol(\Sigma_{A_{12, a}}) = \frac{\pi^3}{4} \int dx\, \hat \rho(\hat x) \left( \hat y_2(\hat x) - \hat y_1(\hat x)  + \frac{2}{3}\right)
   = \frac{21 \pi^3}{64 k} \,, \\
   \Vol(\Sigma_{A_{31, a}}) = \frac{\pi^3}{4} \int dx\, \hat \rho(\hat x) \left( \hat y_1(\hat x) - \hat y_3(\hat x)  + \frac{2}{3}\right)
   = \frac{21 \pi^3}{64 k} \,. 
 }

Let us understand how these volumes are related to the volumes of the divisors computed in \cite{Fabbri:1999hw}.  The cone over $M^{1, 1, 1}$ is a K\"ahler quotient of $\C^5$ by a $U(1)$ that acts with weights $(2, 2, 2, -3, -3)$ on the coordinates $(u_1, u_2, u_3, v_1, v_2)$ parameterizing $\C^5$.  The $\Z_k$ orbifold used to produce the quiver in figure~\ref{C3Z3Figure}  acts by the identification $(v_1, v_2) \sim (v_1 e^{2 \pi i / k}, v_2 e^{-2 \pi i / k})$ leaving the $u_i$ coordinates untouched.  It is natural to identify $A_{23, a}$ with $u_a$, $A_{12, a}$ with $u_a v_1$, and $A_{31, a}$ with $u_a v_2$.  Using the explicit metric on $M^{1, 1, 1}$ the authors of \cite{Fabbri:1999hw} calculated the volumes of the five-cycles corresponding to either $u_a = 0$ or $v_b = 0$ in $M^{1, 1, 1}$ to be
 \es{uvCycles}{
  \Vol(\Sigma_{u_a}) = \frac{3 \pi^3}{16} \,, \qquad
   \Vol(\Sigma_{v_b}) = \frac{9 \pi^3}{64} \,.
 }
We see that these equations are consistent with \eqref{Vols}:  we have $k \Vol(\Sigma_{A_{23, a}}) = \Vol(\Sigma_{u_a})$ as well as $k \Vol(\Sigma_{A_{12, a}}) = \Vol(\Sigma_{u_a}) + \Vol(\Sigma_{v_1})$ and $k \Vol(\Sigma_{A_{31, a}}) = \Vol(\Sigma_{u_a}) + \Vol(\Sigma_{v_2})$.  The factor of $k$ in these formulas comes from the fact that the cycles whose volumes are given in \eqref{Vols} belong to a $\Z_k$ orbifold of $M^{1, 1,1 }$.

For those interested in another simple example of a theory with chiral bifundamental fields, 
we describe our predictions
for a theory with the cone over $Q^{2,2,2}$ as its Abelian moduli space 
in appendix \ref{app:Q222}.

\subsection{Missing operators}

There is a difference between the matrix model and operator counting
that manifests itself in chiral theories.  The matrix model depends
explicitly on the bifundamental fields, and a $\delta y$ saturates
when it reaches minus the R-charge of a bifundamental field.
In the absence of flavors, the saturation of the $\delta y$ is responsible
for all of the corners in $\rho$ and $\rho \delta y$.
In the $\C^3/\Z_3$ example, $\rho$ has a corner at $\hx = \frac{1}{k_+ \Delta_x + \Delta}$.  We might expect that there exists some bifundamental field
$A_{ij,k}$ so that $\delta y + R[A_{ij,k}]$ becomes zero at $\hx = \frac{1}{k_+ \Delta_x + \Delta}$, or equivalently that
$\psi^{(1,1)}_{A_{ij,k}}(r,r\hx)$ becomes zero at $\hx = \frac{1}{k_+ \Delta_x + \Delta}$.  There is no such field.  However, if we consider the density
$\psi^{(1,1)}_{(A_{31,1},A_{31,2},A_{31,3})}$ of operators when we set
$A_{31,1}=A_{31,2}=A_{31,3}=0$, then this density does become zero
at $\hx = \frac{1}{k_+ \Delta_x + \Delta}$.
So it appears to be important to allow arbitrary sets of bifundamental
fields to be set to zero.
A more geometric way of saying this is that the important objects
in the operator counting formula are not the bifundamental fields but rather
five-cycles in the Sasaki-Einstein manifold.  In the $\C^3/\Z_3$ theory,
there seems to be no operator constructed from bifundamental fields
that corresponds to a five-brane wrapping the cycle
$A_{31,1}=A_{31,2}=A_{31,3}=0$.\footnote{%
 Unlike setting $A_{23,1}=A_{23,2}=A_{23,3}=0$ 
 where there are no non-vanishing operators, when we set
 $A_{31,1}=A_{31,2}=A_{31,3}=0$, the number of non-zero operators
 $\psi_{(A_{31,1},A_{31,2},A_{31,3})}$ scales as $r^3$, 
 indicating the presence of a 5-cycle in the geometry.
}
 We might say that we are missing some operators.
We note that the problem could be resolved if we added an operator
$A_{31,1}/A_{23,1}$, since the cycle $A_{31,1}=0$ is the sum of the cycles
$A_{23,1}=0$ and $A_{31,1}=A_{31,2}=A_{31,3}=0$.
The problem never arises in non-chiral non-flavored theories because these
theories do have an operator for every cycle.

The flavored $\mathcal{N}=8$ and flavored ABJM model also have missing
operators.  At $x = 0$, there is a corner in the solutions 
that does not correspond to any $\delta y_{ab}$ saturating at the R-charge of
some bifundamental field $X$.  Instead, the corner comes from the $q$ fields.
From the operator counting perspective,
this corner can be explained by the fact that $\psi^{(1,1)}_{T}$
becomes zero at $\hx = 0$.

\section*{Acknowledgments}

We would like to thank F.~Benini, D.~Jafferis and I.~Klebanov  for discussion.
DG, CH, and SP were supported in part by the NSF under Grant No.~PHY-0756966.
DG and CH were also supported in part by the US NSF under Grant No.~PHY-0844827, and SP by Princeton University through a Porter Ogden Jacobus Fellowship.  CH also thanks the Sloan Foundation for partial support.

\appendix

\section{$F$-Maximization for the necklace quivers}
\label{FMAXIMIZATIONNECKLACE}

We would like to show that to leading order in $N$ the free energy of the necklace quivers with arbitrary R-charges studied in section~\ref{NECKLACE} is maximized when $R[A_a] = R[B_a] = 1/2$ and $\Delta_m = 0$.  
We can only show this if the gauge groups are $SU(N)$.  In the $U(N)^d$ case, the symmetries \eqref{Symmetries} imply that the free energy has flat directions, but we can nevertheless show that the free energy is maximized when the invariant combinations $c_1$ and $c_2$ defined in eq.~\eqref{DeltaDefs} are set to zero.  The critical R-charges correspond to the case where there is ${\cal N} = 3$ supersymmetry as opposed to just ${\cal N} = 2$.

The essential ingredient of the proof is the observation that the polygon ${\cal P}$, which depends on $\vec{c}$, is the polar dual of a polygon ${\cal Q}$ that does not depend on $\vec{c}$ about the unit circle centered at $(-\vec{c}/2)$.   Let $\vec{\beta_a} = (1, q_a)$ and $\vec{c} = (c_1, c_2)$ be vectors in $\R^2$.  The polygon ${\cal Q}$ is the Minkowski sum 
 \es{QDef}{
  {\cal Q} = \left\{\sum_{a=1}^d u_a \vec{\beta}_a \in \R^2: u_a \in (-1/2, 1/2) \right\}
 }
of the vectors $\vec{\beta_a}$.  Indeed, one can rewrite ${\cal P}$ as the intersection of half-planes
 \es{PRewrite}{
  {\cal P} =  \left\{ \vec{t} \in \R^2 :  \frac{1}{2} \vec{t} \cdot \vec{c}   + \sum_{a=1}^d 
   \vec{t} \cdot \left(u_a \vec{\beta}_a \right)
    \leq 1, \forall u_a \in (-1/2, 1/2) \right\} \,.
 }
The boundaries of these half-planes are precisely the polar duals of the points in ${\cal Q}$ about the unit circle centered at $(-\vec{c}/2)$.

Let $\vec{v}_i$ be the vertices of ${\cal Q}$ ordered so that the line segment between $\vec{v}_i$ and $\vec{v}_{i+1}$ is part of the boundary of ${\cal Q}$.  The line passing through $\vec{v}_i$ and $\vec{v}_{i+1}$ is polar dual to a vertex  $\vec{w}_{i, i+1}$ of ${\cal P}$.  Polar duality implies $\vec{w}_{i, i+1} \cdot (\vec{v}_i + \vec{c}/2) = \vec{w}_{i, i+1} \cdot (\vec{v}_{i+1} + \vec{c}/2) = 1$, so
 \es{Gotw}{
 \vec{w}_{i, i+1} =  \frac{*(\vec{v}_{i+1} - \vec{v}_i)}{(*(\vec{v}_{i+1} + \vec{c}/2)) \cdot (\vec{v}_i  + \vec{c}/2)} \,,
 }
where $*$ denotes the Hodge dual in $\R^2$.  By splitting ${\cal P}$ into triangles we can write the area of ${\cal P}$ as
 \es{volPExplicit}{
  \Area({\cal P}) = \sum_i \Area(\vec{w}_{i-1, i}, \vec{w}_{i, i+1}, 0) 
   = \sum_i \frac{1}{2} \abs{\vec{w}_{i-1, i} \cdot (*\vec{w}_{i, i+1}) }\,,
 }
where we denoted the area of a triangle whose vertices are given by the vectors $\vec{\alpha}$, $\vec{\beta}$, and $\vec{\gamma}$ by $\Area (\vec{\alpha}, \vec{\beta}, \vec{\gamma})$.  Using eq.~\eqref{Gotw}, eq.~\eqref{volPExplicit} becomes
 \es{ww}{
  \Area({\cal P}) 
  &= \frac 14 \sum_i \frac{\Area(\vec{v}_{i-1}, \vec{v}_i, \vec{v}_{i+1})}{\Area(\vec{v}_{i}, \vec{v}_{i-1}, -\vec{c}/2 )
    \Area(\vec{v}_{i+1}, \vec{v}_{i}, -\vec{c}/2 )} \,.
 }
As long as $-\vec{c}/2$ belongs to the interior of ${\cal Q}$, 
the Hessian matrix of 
each term in this sum, seen as a function of $\vec{c}$, 
is positive definite, so $\Area({\cal P})$ is a convex function of $\vec{c}$.  
(To compute the Hessian it is easiest to work in a coordinate system where $\vec c$ is parametrized by the distance from two neighboring sides of the polygon to $-\vec c/2$.)

In our case ${\cal Q}$ is symmetric about the origin as can be easily seen from eq.~\eqref{QDef}.  Consequently, $\Area({\cal P})$ is an even function of $\vec{c}$, and we have just shown that it is also convex.   It follows that $\Area({\cal P})$ is minimized for $\vec{c} = 0$.  Equivalently, the free energy is maximized when $\vec{c} = 0$.  Using the $F$-maximization conjecture of \cite{Jafferis:2010un}, we have thus shown that the correct R-charges in the necklace quivers with superpotential \eqref{NecklaceW} satisfy $c_1 = c_2 = 0$.  That's all one can say about the $U(N)^d$ theory.  If the gauge groups are instead $SU(N)$, the tracelessness constraints $\int dx\, \rho(x) \delta y_a(x) = 0$ imply (when $\vec{c} = 0$)
 \es{SUNCondition}{
  \int dx\, \rho(x) \delta y_a(x) = \frac{R[B_a] - R[A_a]}{2} = 0\,,
 }
so $R[A_a] = R[B_a] = 1/2$.  From $c_2 = 0$ we also get $\Delta_m = 0$.

\section{Further examples}
\label{app:furtherexamples}

For notational convenience we set $T^{(1)} = T$ and $T^{(-1)} = \tilde T$.

\subsection{Flavored ABJM theory}
\label{sec:flavoredABJM}

We consider the flavored ABJM model with the superpotential 
 \es{deltaW}{
  W \sim \tr \left[ \epsilon^{ij} \epsilon^{kl} A_i B_k A_j B_l + \sum_{j=1}^{n_{a1}} q_j^{(1)} A_1 \tilde q_j^{(1)} +
   \sum_{j=1}^{n_{a2}} q_j^{(2)} A_2 \tilde q_j^{(2)}
   + \sum_{j=1}^{n_{b1}} Q_j^{(1)} B_1 \tilde Q_j^{(1)} +
   \sum_{j=1}^{n_{b2}} Q_j^{(2)} B_2 \tilde Q_j^{(2)} \right] \,.
 }
When $N = 1$, the superpotential is supplemented by 
 the relation (\ref{TTtildeOPE}) which in this case is 
 $T \tilde T = A_1^{n_{a1}} A_2^{n_{a2}} B_1^{n_{b1}} B_2^{n_{b2}}$  
 \cite{Benini:2009qs, Jafferis:2009th}.
 The corresponding matrix model was solved in the large $N$ limit in \cite{Jafferis:2011zi}.  Our strategy is the same as for the flavored ${\mathcal N}=8$ theory.  In this section, we will review the solution for $\rho(x)$ and $\delta y \equiv y_1 - y_2$.  In the next section, we will compare these results with the distribution of operators in the chiral ring.

 We define $R[A_i] \equiv \Delta_{A_i}$, $R[B_i] \equiv \Delta_{B_i}$, $\Delta \equiv R[T]$,
  and $ \tilde \Delta \equiv R[\tilde T]$. 
  Without loss of generality, we will assume that
 $\Delta_{A_2} < \Delta_{A_1}$ and $\Delta_{B_2} < \Delta_{B_1}$.
To keep the notation concise, we also define
 \es{kpm}{
  k_\pm & \equiv k \pm \frac{1}{2} \left(n_{a1} + n_{a2} - n_{b1} - n_{b2} \right) \,, \\
  \Delta_2 &\equiv \Delta_{A_1} \Delta_{A_2} - \Delta_{B_1} \Delta_{B_2} \ , \\
  \Delta_3 &\equiv  \Delta_{A_1} \Delta_{A_2} (\Delta_{B_1} + \Delta_{B_2})
    + \Delta_{B_1} \Delta_{B_2} (\Delta_{A_1} + \Delta_{A_2}) \ .
 }

Taking the marginality constraints on the R-charges into account, 
in the large $N$ limit, the matrix model free energy functional is
 \es{FreeABJM}{
  \frac{\tilde F[\rho, \delta y]}{2 \pi N^{3/2}} &= 
  \int dx \, \rho \Biggl[
  \frac{1}{2} (k_+ + k_-)  x \, \delta y - \rho \left((\delta y)^2
    + \Delta_2 \, \delta y  - \frac{1}{2} \Delta_3 \right)
    +\frac{1}{2} (\Delta -  \tilde \Delta)  x  
    \\
    &{}
    \hspace{20mm}
        + \frac{1}{2} \abs{x} \left(
     \Delta + \tilde \Delta
   +\left(k_+ - k_- \right)  \delta y 
   \right) 
   \Biggr]
  - \mu \left(\int dx\, \rho -1 \right) \,.
 }

The eigenvalue density has four regions: 
 \begin{eqnarray}
 \label{Left}
  -\frac{1}{R[\tilde T A_2^{k_-}]} <  \hat x   < -\frac{1}{R[\tilde T A_1^{k_-}]}: &&
   \hat \rho = \frac{1 + \hat x R[\tilde T A_2^{k_-}] }{\Delta_3 + 2 \Delta_{A_2} \Delta_2 - 2 \Delta_{A_2}^2} \ , \; \; \;
    \delta \hat y = - \Delta_{A_2} \ ;
  \\
  \label{Middle}
  -\frac{1}{R[\tilde T A_1^{k_-}]} < \hat x  < 0: && 
   \hat \rho  = \frac{2 + 2\hat x  \tilde \Delta
    + \hat x k_- \Delta_2 }
    {\Delta_2^2 + 2 \Delta_3} \ ,
     \nonumber \\
    &&
   \hspace{10mm} \delta \hat y = \frac{k_- \hat x \Delta_3
       -(1 + \hat x \tilde \Delta )  \Delta_2}
    {2 + 2 \hat x  \tilde \Delta
    + \hat x k_- \Delta_2} \ ;
    \\
\label{Middle2}
  0 <  \hat x  < \frac{1}{R[T B_1^{k_+}]}: &&
    \hat \rho  =  \frac{2 - 2 \hat x \Delta
    + \hat x k_+ \Delta_2 }
    {\Delta_2^2 + 2 \Delta_3} \ , \nonumber \\
    &&
     \hspace{10mm} \delta \hat y  = \frac{k_+ \hat x \Delta_3
       -(1 - \hat x \Delta)  \Delta_2}
    {2 - 2 \hat x \Delta
    + \hat x k_+ \Delta_2 } \ ;
\\
\label{Right}
 \frac{1}{R[T B_1^{k_+}]} < \hat x  < \frac{1}{R[T B_2^{k_+}]}:&& 
   \hat \rho = \frac{1 - \hat x R[T B_2^{k_+}]} 
    {\Delta_3- 2 \Delta_{B_2} \Delta_2 - 2 \Delta_{B_2}^2} \ , \; \; \;
    \delta \hat y = \Delta_{B_2} \ ;
 \end{eqnarray}
As in (\ref{hatDefs}), we have introduced the rescaled variables $x = \hat x \mu$ and $\rho(x) = \hat \rho(\hat x) \mu$.

\subsubsection*{Operator counting}

There are operators containing $\tilde T^{-m}$ for $m<0$ and operators containing $T^m$ for $m>0$.  They take the form $T^m A_1^{\alpha_1} A_2^{\alpha_2} B_1^{\beta_1} B_2^{\beta_2}$ and  $\tilde T^{-m} A_1^{\alpha_1} A_2^{\alpha_2} B_1^{\beta_1} B_2^{\beta_2}$, where gauge invariance demands $\alpha_1+ \alpha_2 - \beta_1 - \beta_2 = -m k_\pm$.  If we wanted to count operators that don't vanish when, for example, $A_1=0$, then we just set $\alpha_1=0$. 

We counted the operators using a slightly modified version of the method outlined in 
appendix \ref{app:toric}.  Having written the operators in terms of both $T$ and $\tilde T$, it is simpler to use two different coordinate systems on the cone $C$, one when $m>0$ and one when $m<0$.  The coordinate systems are related by (\ref{TTtildeOPE}).  
The operator counts reproduce (\ref{Left}), (\ref{Middle}), (\ref{Middle2}), and (\ref{Right}) via our conjecture (\ref{results}).

Here are some of the details for the calculation of $\hat \rho(\hat x)$ when $m>0$.
The density of operators is given by
\begin{equation} \label{ABJMOperatorIntegral}
\begin{split}
\frac{\partial^2 \psi}{\partial r \partial m} = &
\int d\alpha_1\,d\alpha_2\,d\beta_1\,d\beta_2\, \delta(\alpha_1+\alpha_2-\beta_1-\beta_2 + mk_+) \\
& \times \delta(r-m \Delta-\alpha_1 \Delta_{A_1} - \alpha_2 \Delta_{A_2} -
\beta_1 \Delta_{B_1} - \beta_2 \Delta_{B_2}) \ .
\end{split}
\end{equation}
This integral gives the area of a slice of a tetrahedron.  The slice is either a triangle or a quadrilateral (which may be regarded as a
triangle with another triangle cut out).
We find for $m \geq 0$
\begin{equation}
\begin{split}
\frac{\partial^2 \psi}{\partial r \partial m} = & 
\sum_{j=1}^2 \frac{ (r-mR[T B_j^{k_+}])^2 \, \theta(r-mR[T B_j^{k_+}]) }{2(\Delta_3- 2 \Delta_{B_j} \Delta_2 - 2 \Delta_{B_j}^2)}
 \ .
\end{split}
\end{equation}
Taking a derivative of this expression with respect to $r$ yields (\ref{Middle2}) and (\ref{Right}).

\subsection{$\C^3/(\Z_2 \times \Z_2)$ theory}
\label{sec:c3z2z2}

\begin {figure} [!t]
  \centering
\newcommand {\svgwidth} {0.45\textwidth}

\begingroup
  \makeatletter
  \providecommand\color[2][]{%
    \errmessage{(Inkscape) Color is used for the text in Inkscape, but the package 'color.sty' is not loaded}
    \renewcommand\color[2][]{}%
  }
  \providecommand\transparent[1]{%
    \errmessage{(Inkscape) Transparency is used (non-zero) for the text in Inkscape, but the package 'transparent.sty' is not loaded}
    \renewcommand\transparent[1]{}%
  }
  \providecommand\rotatebox[2]{#2}
  \ifx\svgwidth\undefined
    \setlength{\unitlength}{217.65859375pt}
  \else
    \setlength{\unitlength}{\svgwidth}
  \fi
  \global\let\svgwidth\undefined
  \makeatother
  \begin{picture}(1,0.76420929)%
    \put(0,0){\includegraphics[width=\unitlength]{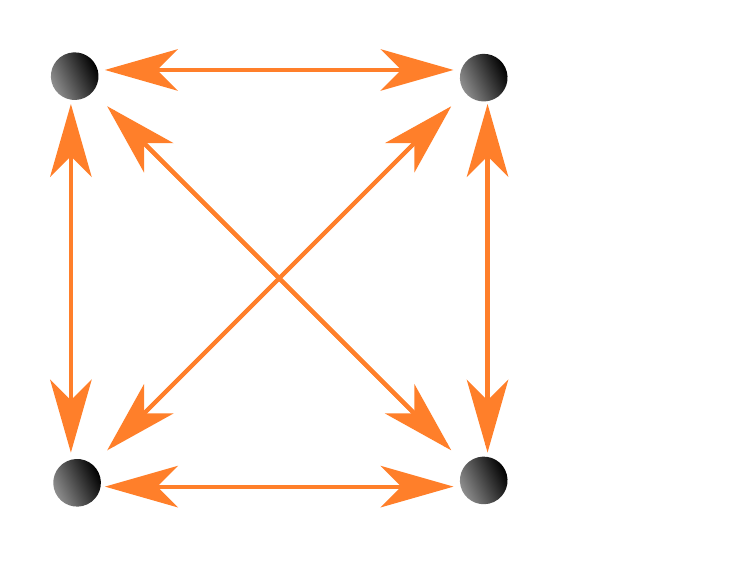}}%
    \put(0.03670561,0.76152463){\color[rgb]{0,0,0}\makebox(0,0)[lt]{\begin{minipage}{0.21527819\unitlength}\raggedright $k_1$\end{minipage}}}%
    \put(0.64320915,0.08431898){\color[rgb]{0,0,0}\makebox(0,0)[lt]{\begin{minipage}{0.28353719\unitlength}\raggedright $k_3$\end{minipage}}}%
    \put(0.17828241,0.71394421){\color[rgb]{0,0,0}\makebox(0,0)[lb]{\smash{$A_{21}$}}}%
    \put(0.64315996,0.7642992){\color[rgb]{0,0,0}\makebox(0,0)[lt]{\begin{minipage}{0.21527819\unitlength}\raggedright $k_2$\end{minipage}}}%
    \put(0.03675482,0.08154439){\color[rgb]{0,0,0}\makebox(0,0)[lt]{\begin{minipage}{0.28353719\unitlength}\raggedright $k_4$\end{minipage}}}%
    \put(0.48520649,0.4722634){\color[rgb]{0,0,0}\makebox(0,0)[lb]{\smash{$A_{42}$}}}%
    \put(0.19116802,0.4722634){\color[rgb]{0,0,0}\makebox(0,0)[lb]{\smash{$A_{31}$}}}%
    \put(0.49069829,0.29126391){\color[rgb]{0,0,0}\makebox(0,0)[lb]{\smash{$A_{13}$}}}%
    \put(0.17828241,0.29126391){\color[rgb]{0,0,0}\makebox(0,0)[lb]{\smash{$A_{24}$}}}%
    \put(-0.00549169,0.2177543){\color[rgb]{0,0,0}\makebox(0,0)[lb]{\smash{$A_{14}$}}}%
    \put(-0.00549169,0.53017017){\color[rgb]{0,0,0}\makebox(0,0)[lb]{\smash{$A_{41}$}}}%
    \put(0.67447233,0.2177543){\color[rgb]{0,0,0}\makebox(0,0)[lb]{\smash{$A_{23}$}}}%
    \put(0.67447233,0.53017017){\color[rgb]{0,0,0}\makebox(0,0)[lb]{\smash{$A_{32}$}}}%
    \put(0.47232089,0.71394421){\color[rgb]{0,0,0}\makebox(0,0)[lb]{\smash{$A_{12}$}}}%
    \put(0.17828241,0.03398025){\color[rgb]{0,0,0}\makebox(0,0)[lb]{\smash{$A_{34}$}}}%
    \put(0.47232089,0.03398025){\color[rgb]{0,0,0}\makebox(0,0)[lb]{\smash{$A_{43}$}}}%
  \end{picture}%
\endgroup

  \caption {The quiver for $\C^3/(\Z_2 \times \Z_2)$.  There are four
  $U(N)$ gauge groups with Chern-Simons coefficients $k_a$.
  The matter content consists of the 12 bifundamental fields $A_{ab}$ for
  $a \ne b$, transforming under the fundamental of the $b$th gauge group
  and the antifundamental of the $a$th gauge group.
  \label{C3Z2Z2Figure}}
\end {figure}
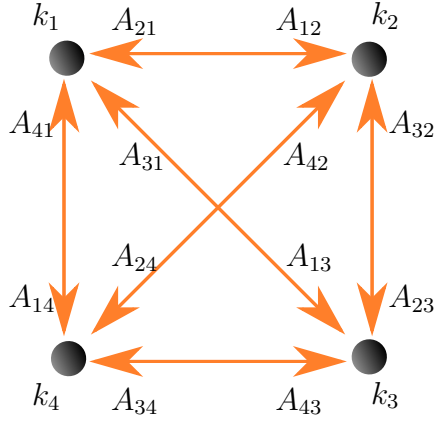

Let's examine the field theory in figure \ref{C3Z2Z2Figure}.  It has four gauge groups with CS levels $k_a$, $a = 1, \ldots, 4$, and twelve bifundamental fields $A_{ab}$ transforming in $(\overline{\bf N}_a, {\bf N}_b)$, one for every ordered pair $(a, b)$ with $a \neq b$.  The superpotential is
 \es{SquareSuperpot}{
  W = \tr \left[ \sum_{a = 1}^4 \epsilon_{abcd} A_{db} A_{cd}  A_{bc} \right] \,.
 }
 The superpotential relations are supplemented by the monopole OPE (\ref{TTtildeOPE}) $T \tilde T = 1$.  We define $R[A_{ab}] \equiv R_{ab}$ and $R[T] = - R[\tilde T] \equiv \Delta$.  
 
The superpotential
 contains eight distinct terms that impose the relations $R_{ab} + R_{bc} + R_{ca}  = 2$ for any triplet $(a, b, c)$ of pairwise distinct gauge groups.  
These eight equations imply the long-range force cancellation \eqref{BetaCond}.
Only seven of these equations are linearly independent, leaving
 five independent R-charges out of the twelve $R_{ab}$. 
 
 Even though for given $k_a$ the matrix model depends on $6$ R-charges ($\Delta$ and the five linearly independent $R_{ab}$), the dependence on three of these parameters is trivial because of the flat directions \eqref{Symmetries}.  We can use these symmetries to reduce the number of independent R-charges to three: $\Delta_x$, $\Delta_y$ and $\Delta$ where
we pick
 \es{RhatChoices}{
   R_{12} = R_{21} = R_{34} =  R_{43} &= \Delta_x \,, \\
   R_{23} = R_{32} =  R_{41} = R_{14} &= \Delta_y \,, \\
   R_{13} =  R_{31} =  R_{24} = R_{42} &= 2 - \Delta_x - \Delta_y \equiv \Delta_z \,. 
 }
The matrix model is then
 \es{MatrixAgain}{
  F[\rho,  y_a] &= 2 \pi N^{3/2} \int dx\, \rho x \sum_{a=1}^d k_a  y_a 
   + 2 \pi \Delta N^{3/2} \int dx\, \rho x \\
   &{}+ \frac{\pi N^{3/2}}{2} \int dx\, \rho^2 \sum_{(a, b, c)} \left(y_b - y_a + R_{ab} \right) 
     \left(y_c - y_b + R_{bc} \right) \left( y_a -  y_c + R_{ca} \right) \ .
 }

For simplicity, let's focus on the case $k_1 = -k_2 = k_3 = -k_4 = k>0$ and take $\Delta_y \geq \Delta_x$.  The saddle point eigenvalue distribution splits into three regions where $\rho$ is linear:
 \es{solnSquare}{
  \frac{1}{\Delta - 2 k \Delta_x} < \hx < \frac{1}{\Delta - 2 k \Delta_y}:& \qquad
   \hr(\hx) = \frac{1 - \hx (\Delta - 2 k \Delta_x)}{4 \Delta_x (\Delta_y - \Delta_x) \Delta_z} \,,
     \qquad \hat y_1 -\hat y_2 = - \Delta_x \,, \\ 
   \frac{1}{\Delta - 2 k \Delta_y} < \hx < \frac{1}{\Delta + 2 k \Delta_y}:& \qquad
   \hr(\hx) = \frac{1 - \hx \Delta}{4 \Delta_x \Delta_y \Delta_z} \,, 
     \qquad \hat y_1 -\hat y_2 = \frac{2k \hx \Delta_x \Delta_y}{1 - \hx \Delta} \,, \\ 
  \frac{1}{\Delta + 2 k \Delta_y} < \hx< \frac{1}{\Delta + 2 k \Delta_x}:& \qquad
   \hr(\hx) = \frac{1 - \hx (\Delta + 2 k \Delta_x)}{4 \Delta_x (\Delta_y - \Delta_x) \Delta_z} \,, 
     \qquad \hat y_1 -\hat y_2 =  \Delta_x \,.
 }
In all three regions, $\hat y_1 = \hat y_3$ and $\hat y_2 = \hat y_4$.

\subsubsection*{Operator counting}

Without the monopole operators, the ring of functions $A_{ab}$ modulo superpotential 
relations is the ring of functions on $\C^3/(\Z_2 \times \Z_2)$.
This ring 
consists of polynomials in $x,y,z$ with the constraint that
the numbers of $x,y,z$ in each term must be either all even or all odd.
We call $A_{12}, A_{21}, A_{34}, A_{43}$ ``$x$ fields'',
$A_{14}, A_{41}, A_{23}, A_{32}$ ``$y$ fields'', and
$A_{13}, A_{31}, A_{24}, A_{42}$ ``$z$ fields''.
We can get a gauge invariant operator by taking a combination of two
$x$ fields (e.g.~$A_{12} A_{21})$, two $y$ fields ($A_{13} A_{31}$),
two $z$ fields ($A_{14} A_{41}$), or one of each type of field
($A_{12} A_{23} A_{31}$).  The gauge invariant operators with $m=0$
are those with an even number of each of $x,y,z$, or an odd number of each
of $x,y,z$.

Adding back the monopole operators yields a ring of functions on a four-dimensional
cone.
An electric charge of $(1,-1,1,-1)$ from $T$ can be cancelled out by two $x$'s
($A_{12} A_{34}$) or two $y$'s ($A_{14} A_{32})$, but not by $z$'s.
So, if we have an operator of the schematic form $T^m x^{n_x} y^{n_y} z^{n_z}$ for $m>0$
and $\tilde T^{-m} x^{n_x} y^{n_y} z^{n_z}$ for $m<0$, then we have the
constraint $n_x + n_y \ge 2|m|$.
The operator density is then
\begin{equation}
\frac{\partial^2 \psi}{\partial r \partial m}= \frac{1}{4} \int dn_x\,dn_y\,dn_z\,
\theta(n_x+n_y-2k|m|) \delta(r-\Delta_x n_x - \Delta_y n_y - \Delta_z n_z - \Delta m) \ .
\end{equation}
The factor of $\frac{1}{4}$ comes from the constraint that the numbers
of $x,y,z$ must be all even or all odd.

\begin{figure}
a) \includegraphics[width=2.5in]{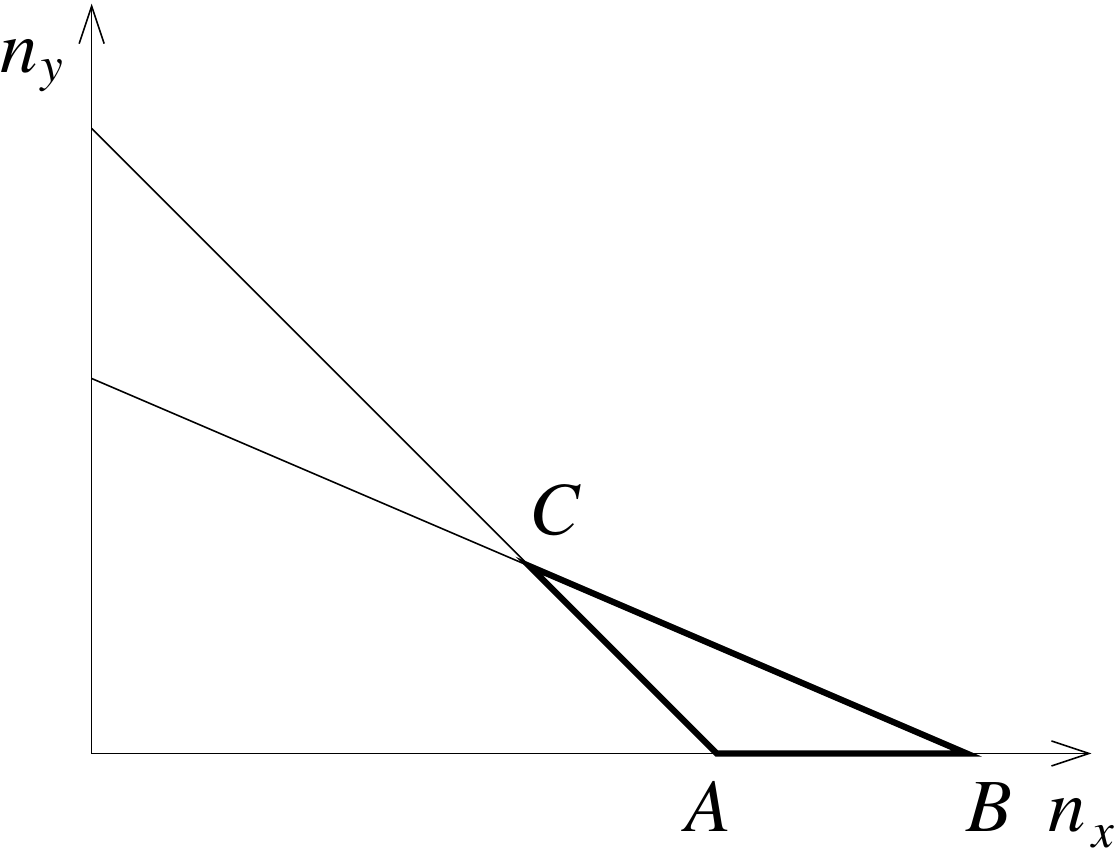}
b) \includegraphics[width=2.5in]{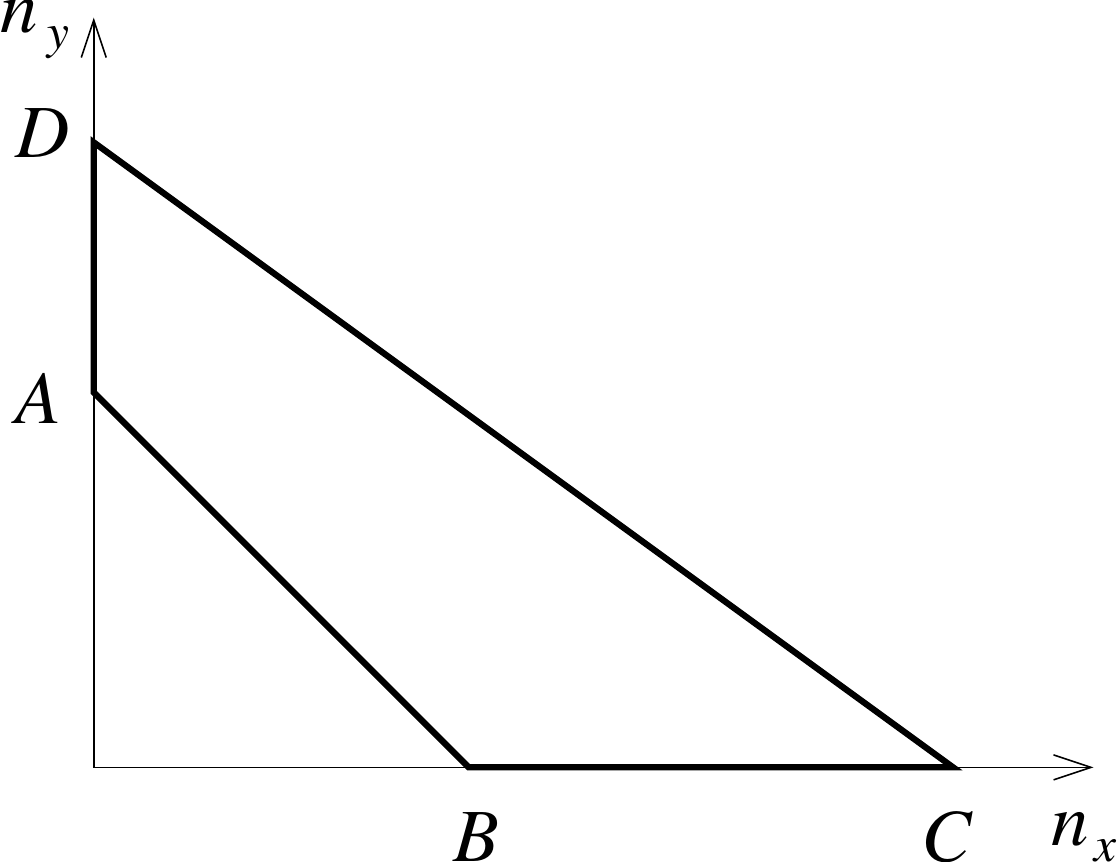}
\caption{The area of the polygonal regions $ABC$ and $ABCD$ is proportional to 
$\partial^2 \psi / \partial r \partial m$ for the
$\mathbb{C}^3 / \mathbb{Z}_2 \times \mathbb{Z}_2$ quiver: a) $r - \Delta m - 2 k |m| \Delta_y <0$;
b) $r - \Delta m - 2 k |m| \Delta_y >0$
\label{fig:z2z2tritrap}
}
\end{figure}

Performing the integral over $n_z$ introduces an overall factor of $1/\Delta_z$.
The remaining integral reduces to the area of a polygonal region satisfying the constraints
$n_y>0$, $n_x>0$, $n_x + n_y > 2 k |m|$, and $\Delta_x n_x + \Delta_y n_y < r - \Delta m$.
For small $|m|$, the polygonal region is a quadrilateral while for large $|m|$, the region is a triangle (see figure \ref{fig:z2z2tritrap}).
Assuming that $\Delta_y > \Delta_x$, we find
\begin{equation}
\frac{\partial^2 \psi}{\partial r \partial m} =
\begin{cases}
\frac{1}{8\Delta_z} \frac{(r - \Delta m - 2 k |m| \Delta_x)^2}{\Delta_x (\Delta_y- \Delta_x)^2} &
\mbox{if    } r - \Delta m - 2 k |m| \Delta_y < 0 \ , \\
\frac{1}{8\Delta_z} \left[ \frac{(r-\Delta m)^2}{\Delta_x \Delta_y} - (2 k |m|)^2 \right] & \mbox{if    }
r - \Delta m - 2 k |m| \Delta_y > 0 \ .
\end{cases}
\end{equation}
Taking an additional derivative with respect to $r$, we can easily check that this formula agrees with (\ref{solnSquare}).

Now, in order to compute $\hat y_1(\hx)-\hat y_4(\hx)$,
we count gauge invariant operators with $A_{14}$ set to zero. 
Because of the superpotential relations, all operators with a $z$ are set to zero.
The factor of 1/4 remains the same because now we may only consider operators with even numbers of $x$ and $y$ fields.
The expression for $\partial \psi_{14} / \partial m$ is given by the area of the same polygonal region that governs $\partial^2 \psi / \partial r \partial m$, 
but we lose the factor of $\Delta_z$ because we drop the integral over $n_z$:
\begin{equation}
\frac{\partial^2 \psi_{14}}{\partial r \partial m} = \Delta_z \frac{\partial^3 \psi}{\partial r^2 \partial m} \ .
\end{equation}
Therefore, we have $\hr(\hx) (\Delta_z + \hat y_1(\hx)-\hat y_4(\hx)) = \hr(\hx) \Delta_z$, and hence $\hat y_1(\hx)=\hat y_4(\hx)$.  A similar calculation shows
$\hat y_2(\hx)=\hat y_3(\hx)$.

Finally we count the operators with $A_{12}$ set to zero.
Most operators with an $x$ will become zero.  However, fields containing
only $T$, $A_{21}$, $A_{43}$, $A_{23}$, $A_{41}$, and the $z$ fields
are not set to zero by the superpotential relations.
So the nonzero fields are those with $n_x=0$ and an even number of $y$ and $z$ fields, or $m \ge 0$ and $n_x+n_y=2km$ with an even number of $z$ fields.
After a little work, we find
\begin{equation}
\frac{\partial^2 \psi_{12}}{\partial r \partial m} = 
\begin{cases}
0 & \mbox{if    } m/r < -(2 k \Delta_y - \Delta)^{-1} \\
\frac{r - \Delta m + 2 k m \Delta_y}{4 \Delta_y  \Delta_z} & \mbox{if    } -(2 k \Delta_y - \Delta)^{-1} < m/r < (2 k \Delta_y + \Delta)^{-1} \\
\frac{r - \Delta m - 2 k m \Delta_x}{2 \Delta_z (\Delta_y - \Delta_x)} & \mbox{if     } m/r > (2 k \Delta_y + \Delta)^{-1}
\end{cases}
\end{equation}
This result matches $\hat y_1(\hat x) - \hat y_2(\hat x)$ computed from (\ref{solnSquare}).

\section{Toric varieties in general}
\label{app:toric}

By toric moduli space we mean more specifically that the moduli space for the Abelian gauge theory is an eight-dimensional toric Calabi-Yau cone $V$.  That $V$ is toric means it is a $T^4$ 
torus fibration over a four-dimensional rational polyhedral cone $C$.  This polyhedral cone
is the set of points satisfying
\begin{equation}
C = \{ y \in \mathbb{R}^4 :  y \cdot  v_a \geq 0 \} \ ,
\end{equation}
where $v_a \in \mathbb{Z}^4$, $a = 1, \ldots, n$, are inward pointing vectors normal to the faces $F_a$ of the cone:
\begin{equation}
F_a = \{  y \in C : y \cdot v_a = 0 \} \ .
\end{equation} 
The fact that $V$ is Calabi-Yau implies that the end-points of the vectors $v_a$ lie in a common hyperplane $\mathbb{R}^3$.  

One convenient aspect of this construction is that lattice points in $C$ correspond to operators in the chiral ring of the Chern-Simons theory.  The coordinates of a lattice point are the $U(1)$ global charges of the operator.  The vector $b$ 
that measures the R-charge is often called the Reeb vector where the R-charge is then $r = y \cdot b$.  The vectors $v_a$ 
correspond to other global charges, $q_a = y \cdot v_a$, and we can introduce additional charges as well.  In the gauge theories considered in this paper, the monopole charge $m$ played an important role.  Let us introduce $ t$ as the vector that measures monopole charge.

We introduced previously the function $\psi(r,m)$ as the number of operators with R-charge less than $r$ and monopole charge less than $m$.  From the toric perspective, this function in the large $r$ and $m$ limit is the volume of a four-dimensional polytope:
\begin{equation}
C_{r,m} = C \cap \{  y \cdot  b \leq r \} \cap \{  y \cdot  t \leq m \} \ ,
\end{equation}
where $\psi(r,m) = \Vol(C_{r,m})$.

We would like to understand geometrically how to compute derivatives of $\psi(r,m)$.  The value of $\psi(r,m)$ is a four-dimensional integral we can write as
\begin{equation}
\psi(r,m) = \int_{C_{r,m}} d^4 y \ .
\end{equation}
To take a derivative of $\psi$ with respect to $r$, we can rotate the coordinate system so that one of the $y$'s points in the direction of $b$ and replace $d^4y$ with $d^3 y \, dr / |b|$ where $|b|$ is the Jacobian factor from the change of variables.  The derivative is then related to the three-dimensional volume of the polyhedron
\begin{equation}
D_{r,m} = C \cap \{  y \cdot  b = r \} \cap \{  y \cdot  t \leq m \} \ 
\end{equation}
where
$
\partial \psi / \partial r =\Vol(D_{r,m})/ |b|
$.\footnote{%
 This last expression may seem strange because the right hand 
 side seems to depend on a metric while the left hand side depends 
 only on a volume form on $C$.  Interpreting $\Vol(D_{r,m})$ as a three 
 form instead of a number, we could rewrite this expression in a manifestly
 metric independent way:
 $(\partial \psi / \partial r) t = \star \Vol(D_{r,m})$. 
} 

Similarly,
we can visualize $\partial^2 \psi / \partial r \partial m$ 
as the area of a two-dimensional polygon $P_{r,m}$:
\begin{equation}
P_{r,m} = C \cap \{  y \cdot b = r \} \cap \{   y \cdot t = m \} \ .
\end{equation}
Now we rotate our coordinate system so that two of the $y$'s lie in the plane spanned by $b$ and $t$.  The Jacobian factor is $|t \wedge b | = \sqrt{t^2 b^2 - ( t \cdot  b)^2}$.  Geometrically, the second partial is
\begin{equation}
\frac{\partial^2 \psi}{\partial r \partial m} = \frac{\operatorname{Area}(P_{r,m})}{|t \wedge b | } \ .
\label{toricd2psi}
\end{equation}

The function $\psi_X(r,m)$ has a toric interpretation as well.  In the examples we considered, $X$ corresponds to an integer linear combination of the $v_a$.  Let us consider the simple case where $X_a$ corresponds to a single $v_a$.  
Operators with no $X_a$ are contained in the face $F_a \subset C$.
This fact suggests a relation between $\psi_{X_a} (r,m)$ and a generalization of $\psi(r,m)$ involving a third charge $q_a$, $\psi(r,m,q_a)$.  In particular, it is true that
\begin{equation}
\psi_{X_a}(r,m) = \psi^{(0,0,1)}(r,m,0)  \ .
\end{equation}
 Operators with no $X_a$
and fixed $m$ and $r$ lie along a line $L_{a,m,r} \subset F_a$:
\begin{equation}
L_{a,m,r} = F_a \cap \{ y \cdot b = r \} \cap \{y \cdot t = m \} \ .
\end{equation}
Generalizing the argument used to derive (\ref{toricd2psi}) 
to one more charge, we find
\begin{equation}
\label{toricd3psi}
\frac{\partial \psi_{X_a}^2}{\partial r \partial m} =
\psi^{(1,1,1)}(r,m,0) = 
 \frac{\operatorname{Length}(L_{a,m,r})}
{|t \wedge b \wedge v_a | } 
\ .
\end{equation}
Eqs.\ (\ref{toricd2psi}) and (\ref{toricd3psi}) provide a convenient starting point for counting chiral operators in the examples in the text.

\section{The Cone over $Q^{2, 2, 2}/\Z_k$}
\label{app:Q222}

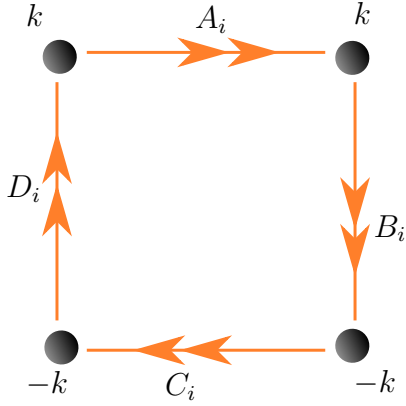
\begin {figure} [!t]
  \centering
\newcommand {\svgwidth} {0.37\textwidth}

\begingroup
  \makeatletter
  \providecommand\color[2][]{%
    \errmessage{(Inkscape) Color is used for the text in Inkscape, but the package 'color.sty' is not loaded}
    \renewcommand\color[2][]{}%
  }
  \providecommand\transparent[1]{%
    \errmessage{(Inkscape) Transparency is used (non-zero) for the text in Inkscape, but the package 'transparent.sty' is not loaded}
    \renewcommand\transparent[1]{}%
  }
  \providecommand\rotatebox[2]{#2}
  \ifx\svgwidth\undefined
    \setlength{\unitlength}{185.00548096pt}
  \else
    \setlength{\unitlength}{\svgwidth}
  \fi
  \global\let\svgwidth\undefined
  \makeatother
  \begin{picture}(1,0.895438)%
    \put(0,0){\includegraphics[width=\unitlength]{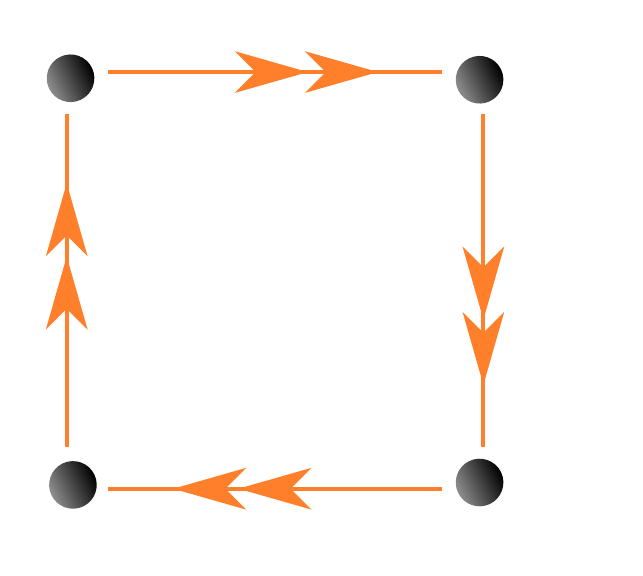}}%
    \put(0.03672311,0.88901527){\color[rgb]{0,0,0}\makebox(0,0)[lt]{\begin{minipage}{0.25327437\unitlength}\raggedright $k$\end{minipage}}}%
    \put(0.75027337,0.09228411){\color[rgb]{0,0,0}\makebox(0,0)[lt]{\begin{minipage}{0.33358096\unitlength}\raggedright $-k$\end{minipage}}}%
    \put(0.75021548,0.89227953){\color[rgb]{0,0,0}\makebox(0,0)[lt]{\begin{minipage}{0.25327437\unitlength}\raggedright $k$\end{minipage}}}%
    \put(0.03678101,0.08901982){\color[rgb]{0,0,0}\makebox(0,0)[lt]{\begin{minipage}{0.33358096\unitlength}\raggedright $-k$\end{minipage}}}%
    \put(-0.00646097,0.46874458){\color[rgb]{0,0,0}\makebox(0,0)[lb]{\smash{$D_i$}}}%
    \put(0.79351539,0.38226065){\color[rgb]{0,0,0}\makebox(0,0)[lb]{\smash{$B_i$}}}%
    \put(0.40433773,0.83630126){\color[rgb]{0,0,0}\makebox(0,0)[lb]{\smash{$A_i$}}}%
    \put(0.33947479,0.03632496){\color[rgb]{0,0,0}\makebox(0,0)[lb]{\smash{$C_i$}}}%
  \end{picture}%
\endgroup

  \caption {Quiver gauge theory believed to be dual to $AdS_4 \times Q^{2, 2,2}/\Z_k$.
  \label{Q222Figure}}
\end {figure}

As another example with chiral bifundamental fields, we can examine the square quiver in figure~\ref{Q222Figure} with CS levels $(k, k, -k, -k)$ and matter fields $A_i$, $B_i$, $C_i$, and $D_i$, with $i = 1, 2$.  With the superpotential is
 \es{WQ222}{
  W \sim \tr \left[ \epsilon^{ij} \epsilon^{kl} D_i C_k B_j A_l \right] 
 }
this quiver is thought to be dual to $AdS_4 \times Q^{2, 2, 2} / \Z_k$ \cite{Hanany:2008fj, Franco:2009sp}.  The quiver has two flavor $SU(2)$ symmetries, one under which $A_i$ and $C_i$ transform as doublets, and one under which $B_i$ and $D_i$ transform as doublets, so one expects the R-charges of the fields belonging to the same edge of the quiver to be equal when $F$ is maximized.  Using the flat directions \eqref{Symmetries} and taking into account the marginality of the superpotential \eqref{WQ222}, one can then set the R-charges of all the bifundamental fields equal to $1/2$ and $\Delta_m = 0$.  With this choice one can go through the operator counting exercise in the Abelian theory and predict that
 \es{MatrixModelPredQ222}{
  \hat \rho(\hat x) &=  \theta\left(\frac{1}{2k} - \abs{\hat x} \right)  
     + \frac{1}{4k} \delta \left(\frac{1}{2k} + \hat x \right)
     +  \frac{1}{4k} \delta \left(\frac{1}{2k} - \hat x \right) \,, \\
  \hat \rho(\hat x) \left(\hat y_2(\hat x) - \hat y_1(\hat x) \right) 
    &= - \frac{1}{8k} \delta \left(\frac{1}{2k} + \hat x \right)
      -  \frac{1}{8k} \delta \left(\frac{1}{2k} - \hat x \right)  \,, \\
  \hat \rho(\hat x) \left(\hat y_3(\hat x) - \hat y_2(\hat x) \right) 
    &= \frac{3}{8k} \delta \left(\frac{1}{2k} + \hat x \right)
      -  \frac{1}{8k} \delta \left(\frac{1}{2k} - \hat x \right)  \,, \\
  \hat \rho(\hat x) \left(\hat y_4(\hat x) - \hat y_3(\hat x) \right) 
    &= - \frac{1}{8k} \delta \left(\frac{1}{2k} + \hat x \right)
      -  \frac{1}{8k} \delta \left(\frac{1}{2k} - \hat x \right)  \,.  
 }

As a consistency check, one can compute the volumes 
 \es{VolumesQ222}{
  \Vol(Y) &= \frac{\pi^4}{24} \int d\hat x\,  \hat \rho(\hat x) = \frac{\pi^4}{16k} \,, \\
  \Vol(\Sigma_{A_i}) &= \frac{\pi^3}{4} \int d\hat x\, \hat \rho(\hat x) 
    \left( \hat y_2(\hat x) - \hat y_1(\hat x)  + \frac{1}{2}\right) = \frac{\pi^3}{8k} \,, \\
  \Vol(\Sigma_{B_i}) &= \frac{\pi^3}{4} \int d\hat x\, \hat \rho(\hat x) 
    \left( \hat y_3(\hat x) - \hat y_2(\hat x)  + \frac{1}{2}\right) = \frac{\pi^3}{4k} \,, \\
  \Vol(\Sigma_{C_i}) &= \frac{\pi^3}{4} \int d\hat x\, \hat \rho(\hat x)
    \left( \hat y_4(\hat x) - \hat y_3(\hat x)  + \frac{1}{2}\right) = \frac{\pi^3}{8k} \,, \\
  \Vol(\Sigma_{D_i}) &= \frac{\pi^3}{4} \int d\hat x\, \hat \rho(\hat x)
    \left( \hat y_1(\hat x) - \hat y_4(\hat x)  + \frac{1}{2}\right) = \frac{\pi^3}{4k} \,. \\
 }
Since $\Vol(Q^{2, 2,2}) = \pi^4 / 16$ \cite{Fabbri:1999hw}, we see that $\Vol(Y)$ matches that of a $\Z_k$ orbifold of $Q^{2, 2,2}$.  As for $M^{1, 1,1}$, we can relate the volumes of the five-cycles in \eqref{VolumesQ222} to those computed in \cite{Fabbri:1999hw}.  The cone over $Q^{2, 2, 2}$ is a $U(1)^2$ K\"ahler quotient of $\C^6$ with weights $(1, 1, -1, -1, 0, 0)$ and $(1, 1, 0, 0, -1, -1)$, together with a $\Z_2$ quotient that flips the sign of $(a_1, a_2)$.  If we denote the coordinates in $\C^6$ by $(a_1, a_2, b_1, b_2, c_1, c_2)$, we have \cite{Fabbri:1999hw}
 \es{VolumesQ222Geometry}{
  \Vol(Q^{2, 2, 2}) &= \frac{\pi^4}{16} \,, \qquad  \Vol(\Sigma_{a_i}) 
    = \Vol(\Sigma_{b_i}) = \Vol(\Sigma_{c_i}) = \frac{\pi^3}{8} \,.
 }
One can think of the $\Z_k$ orbifold as acting on $c_i$ with opposite phases, so it is natural to interpret $A_i$ and $C_i$ as correponding to $a_i$, $B_i$ as corresponding to $b_i c_1$, and $D_i$ as corresponding to $b_i c_2$.   Indeed $k \Vol(\Sigma_{A_i}) = \Vol(\Sigma_{C_i}) = \Vol(\Sigma_{a_i})$, $k \Vol(\Sigma_{B_i}) = \Vol(\Sigma_{b_i}) + \Vol(\Sigma_{c_1})$, and $k \Vol(\Sigma_{D_i}) = \Vol(\Sigma_{b_i}) + \Vol(\Sigma_{c_2})$, the factor of $k$ appearing because the volumes \eqref{VolumesQ222} are computed in a $\Z_k$ orbifold of $Q^{2, 2,2}$.

\bibliographystyle{ssg}
\bibliography{necklace}

\end{document}